\documentclass[sigconf]{acmart}

\newcommand{\MSnPrefixesCloudflare}{{1\,698}}

\newcommand{\MSnPrefixesAmazonWS}{{7\,483}}

\newcommand{\MSnPrefixesAkamai}{{5\,056}}

\newcommand{\MSnPrefixesDDoSGuard}{{25}}

\newcommand{\MSnCDNPrefixesTotal}{{14\,262}}

\newcommand{\CSnDDoSAttacksRounded}{{1.7M}}
\newcommand{\CSnTelegramMessages}{{441}}
\newcommand{\CSnTelegramReplies}{{57\,757}}
\newcommand{\CSnTelegramRepliesRounded}{{58k}}

\newcommand{\CSnTelegramEmojisRounded}{{900k}}
\newcommand{\CSnDefacementRawTotal}{{357\,848}}
\newcommand{\CSnDefacementRawTotalRounded}{{358k}}
\newcommand{\CSnDefacementRawMinusUnifiedTotal}{{40\,799}}
\newcommand{\CSnDefacementRawMinusUnifiedProportionTotal}{{11.00}}
\newcommand{\CSnDefacementUnifiedTotal}{{317\,049}}

\newcommand{\CSnDefacementUnifiedValidTotal}{{274\,963}}
\newcommand{\CSnDefacementUnifiedValidProportionTotal}{{86.73}}
\newcommand{\CSnDefacementUnifiedInvalidTotal}{{42\,086}}
\newcommand{\CSnDefacementUnifiedInvalidProportionTotal}{{13.27}}
\newcommand{\CSnDefacementUnifiedValidStaffVerificationTotal}{{137\,339}}
\newcommand{\CSnDefacementUnifiedValidAutoVerificationTotal}{{97\,652}}
\newcommand{\CSnDefacementUnifiedValidManualVerificationTotal}{{39\,972}}
\newcommand{\CSnDefacersRawTotal}{{4\,347}}

\newcommand{\CSnValidDefacersTotal}{{2\,781}}
\newcommand{\CSnValidDefacersProportionTotal}{{77.44}}
\newcommand{\CSnInvalidDefacersTotal}{{1\,656}}
\newcommand{\CSnInvalidDefacersProportionTotal}{{46.12}}
\newcommand{\CSnTotalHandlesTotal}{{3\,454}}

\newcommand{\CSnDefacementRawZoneH}{{164\,312}}

\newcommand{\CSnDefacementUnifiedZoneH}{{164\,312}}

\newcommand{\CSnDefacementUnifiedValidZoneH}{{143\,485}}
\newcommand{\CSnDefacementUnifiedValidProportionZoneH}{{87.32}}
\newcommand{\CSnDefacementUnifiedInvalidZoneH}{{20\,827}}
\newcommand{\CSnDefacementUnifiedInvalidProportionZoneH}{{12.68}}

\newcommand{\CSnDefacersRawZoneH}{{2\,173}}

\newcommand{\CSnValidDefacersZoneH}{{1\,655}}
\newcommand{\CSnValidDefacersProportionZoneH}{{82.01}}
\newcommand{\CSnInvalidDefacersZoneH}{{843}}
\newcommand{\CSnInvalidDefacersProportionZoneH}{{41.77}}
\newcommand{\CSnTotalHandlesZoneH}{{1\,790}}

\newcommand{\CSnDefacementRawZoneXSec}{{53\,852}}

\newcommand{\CSnDefacementUnifiedZoneXSec}{{53\,814}}

\newcommand{\CSnDefacementUnifiedValidZoneXSec}{{53\,705}}
\newcommand{\CSnDefacementUnifiedValidProportionZoneXSec}{{99.80}}
\newcommand{\CSnDefacementUnifiedInvalidZoneXSec}{{109}}
\newcommand{\CSnDefacementUnifiedInvalidProportionZoneXSec}{{0.20}}

\newcommand{\CSnDefacersRawZoneXSec}{{561}}

\newcommand{\CSnValidDefacersZoneXSec}{{541}}
\newcommand{\CSnValidDefacersProportionZoneXSec}{{99.82}}
\newcommand{\CSnInvalidDefacersZoneXSec}{{24}}
\newcommand{\CSnInvalidDefacersProportionZoneXSec}{{4.43}}
\newcommand{\CSnTotalHandlesZoneXSec}{{560}}

\newcommand{\CSnDefacementRawHaxorID}{{34\,482}}

\newcommand{\CSnDefacementUnifiedHaxorID}{{34\,465}}

\newcommand{\CSnDefacementUnifiedValidHaxorID}{{34\,439}}
\newcommand{\CSnDefacementUnifiedValidProportionHaxorID}{{99.92}}
\newcommand{\CSnDefacementUnifiedInvalidHaxorID}{{26}}
\newcommand{\CSnDefacementUnifiedInvalidProportionHaxorID}{{0.08}}

\newcommand{\CSnDefacersRawHaxorID}{{484}}

\newcommand{\CSnValidDefacersHaxorID}{{443}}
\newcommand{\CSnValidDefacersProportionHaxorID}{{99.55}}
\newcommand{\CSnInvalidDefacersHaxorID}{{15}}
\newcommand{\CSnInvalidDefacersProportionHaxorID}{{3.37}}
\newcommand{\CSnTotalHandlesHaxorID}{{482}}

\newcommand{\CSnDefacementRawDefacerPro}{{28\,594}}

\newcommand{\CSnDefacementUnifiedDefacerPro}{{27\,662}}

\newcommand{\CSnDefacementUnifiedValidDefacerPro}{{26\,379}}
\newcommand{\CSnDefacementUnifiedValidProportionDefacerPro}{{95.36}}
\newcommand{\CSnDefacementUnifiedInvalidDefacerPro}{{1\,283}}
\newcommand{\CSnDefacementUnifiedInvalidProportionDefacerPro}{{4.64}}

\newcommand{\CSnDefacersRawDefacerPro}{{540}}

\newcommand{\CSnValidDefacersDefacerPro}{{486}}
\newcommand{\CSnValidDefacersProportionDefacerPro}{{97.79}}
\newcommand{\CSnInvalidDefacersDefacerPro}{{147}}
\newcommand{\CSnInvalidDefacersProportionDefacerPro}{{29.58}}
\newcommand{\CSnTotalHandlesDefacerPro}{{526}}

\newcommand{\CSnDefacementRawOwnzYou}{{76\,608}}

\newcommand{\CSnDefacementUnifiedOwnzYou}{{67\,510}}

\newcommand{\CSnDefacementUnifiedValidOwnzYou}{{47\,657}}
\newcommand{\CSnDefacementUnifiedValidProportionOwnzYou}{{70.59}}
\newcommand{\CSnDefacementUnifiedInvalidOwnzYou}{{19\,853}}
\newcommand{\CSnDefacementUnifiedInvalidProportionOwnzYou}{{29.41}}

\newcommand{\CSnDefacersRawOwnzYou}{{1\,214}}

\newcommand{\CSnValidDefacersOwnzYou}{{553}}
\newcommand{\CSnValidDefacersProportionOwnzYou}{{54.00}}
\newcommand{\CSnInvalidDefacersOwnzYou}{{722}}
\newcommand{\CSnInvalidDefacersProportionOwnzYou}{{70.51}}
\newcommand{\CSnTotalHandlesOwnzYou}{{689}}

\newcommand{\CSdefacementProportionUSAfterWar}{{21.97}}

\newcommand{\CSdefacementProportionRUBeforeWar}{{0.60}}
\newcommand{\CSdefacementRUAfterWar}{{209}}
\newcommand{\CSdefacementProportionRUAfterWar}{{14.48}}

\newcommand{\CSdefacementProportionUAOneDayAfterWar}{{0.47}}
\newcommand{\CSdefacementUATwoDayAfterWar}{{69}}
\newcommand{\CSdefacementProportionUATwoDayAfterWar}{{6.30}}
\newcommand{\CStotalDefacementProportionTopCountries}{{69.85}}
\newcommand{\CStotalDefacementProportionUS}{{26.95}}
\newcommand{\CStotalDefacementProportionIN}{{11.47}}
\newcommand{\CStotalDefacementProportionID}{{8.41}}
\newcommand{\CStotalDefacementProportionRU}{{1.58}}
\newcommand{\CStotalDefacementProportionUA}{{0.57}}
\newcommand{\CStotalDDoSVictimProportionTopCountries}{{70.49}}
\newcommand{\CStotalDDoSVictimProportionUS}{{24.68}}
\newcommand{\CStotalDDoSVictimProportionBR}{{11.99}}
\newcommand{\CStotalDDoSVictimProportionBD}{{8.10}}
\newcommand{\CStotalDDoSVictimProportionRU}{{3.61}}
\newcommand{\CStotalDDoSVictimProportionUA}{{1.57}}
\newcommand{\CSnDefacementHostedOnCDN}{{4.87}}

\newcommand{\CSnDefacementsRUSignificanceTestDoF}{{2}}
\newcommand{\CSnDefacementsRUSignificanceTestStat}{{12.24}}
\newcommand{\CSnDefacementsRUSignificanceTestPValue}{{p < .01}}

\newcommand{\CSnDefacementsRUEffectSize}{{0.06}}

\newcommand{\CSnDefacementsRUPostHocPValueONETWO}{{p < .05}}

\newcommand{\CSnDefacementsRUPostHocPValueONETHREE}{{p = .3544}}

\newcommand{\CSnDefacementsRUPostHocPValueTWOTHREE}{{p < .001}}

\newcommand{\CSnDefacementsUASignificanceTestDoF}{{2}}
\newcommand{\CSnDefacementsUASignificanceTestStat}{{17.86}}
\newcommand{\CSnDefacementsUASignificanceTestPValue}{{p < .001}}

\newcommand{\CSnDefacementsUAEffectSize}{{0.09}}

\newcommand{\CSnDefacementsUAPostHocPValueONETWO}{{p < .001}}

\newcommand{\CSnDefacementsUAPostHocPValueONETHREE}{{p = .8377}}

\newcommand{\CSnDefacementsUAPostHocPValueTWOTHREE}{{p < .0001}}

\newcommand{\CSnDefacementsUSSignificanceTestDoF}{{2}}
\newcommand{\CSnDefacementsUSSignificanceTestStat}{{17.84}}
\newcommand{\CSnDefacementsUSSignificanceTestPValue}{{p < .001}}

\newcommand{\CSnDefacementsUSEffectSize}{{0.09}}

\newcommand{\CSnDefacementsUSPostHocPValueONETWO}{{p < .01}}

\newcommand{\CSnDefacementsUSPostHocPValueONETHREE}{{p = .1435}}

\newcommand{\CSnDefacementsUSPostHocPValueTWOTHREE}{{p < .0001}}

\newcommand{\CSnDefacementsBRSignificanceTestDoF}{{2}}
\newcommand{\CSnDefacementsBRSignificanceTestStat}{{3.60}}
\newcommand{\CSnDefacementsBRSignificanceTestPValue}{{p = .1656}}

\newcommand{\CSnDefacementsBREffectSize}{{0.01}}

\newcommand{\CSnDefacementsBRPostHocPValueONETWO}{{p = .6481}}

\newcommand{\CSnDefacementsBRPostHocPValueONETHREE}{{p = .1725}}

\newcommand{\CSnDefacementsBRPostHocPValueTWOTHREE}{{p = .0912}}

\newcommand{\CSnDefacementsDESignificanceTestDoF}{{2}}
\newcommand{\CSnDefacementsDESignificanceTestStat}{{3.43}}
\newcommand{\CSnDefacementsDESignificanceTestPValue}{{p = .1796}}

\newcommand{\CSnDefacementsDEEffectSize}{{0.01}}

\newcommand{\CSnDefacementsDEPostHocPValueONETWO}{{p = .2339}}

\newcommand{\CSnDefacementsDEPostHocPValueONETHREE}{{p = .5269}}

\newcommand{\CSnDefacementsDEPostHocPValueTWOTHREE}{{p = .0639}}

\newcommand{\CSnDefacementsINSignificanceTestDoF}{{2}}
\newcommand{\CSnDefacementsINSignificanceTestStat}{{4.21}}
\newcommand{\CSnDefacementsINSignificanceTestPValue}{{p = .1221}}

\newcommand{\CSnDefacementsINEffectSize}{{0.01}}

\newcommand{\CSnDefacementsINPostHocPValueONETWO}{{p = .9049}}

\newcommand{\CSnDefacementsINPostHocPValueONETHREE}{{p = .0670}}

\newcommand{\CSnDefacementsINPostHocPValueTWOTHREE}{{p = .1423}}

\newcommand{\CSnDefacementsIDSignificanceTestDoF}{{2}}
\newcommand{\CSnDefacementsIDSignificanceTestStat}{{10.90}}
\newcommand{\CSnDefacementsIDSignificanceTestPValue}{{p < .01}}

\newcommand{\CSnDefacementsIDEffectSize}{{0.05}}

\newcommand{\CSnDefacementsIDPostHocPValueONETWO}{{p < .01}}

\newcommand{\CSnDefacementsIDPostHocPValueONETHREE}{{p = .1566}}

\newcommand{\CSnDefacementsIDPostHocPValueTWOTHREE}{{p < .05}}

\newcommand{\CSnDefacementsCASignificanceTestDoF}{{2}}
\newcommand{\CSnDefacementsCASignificanceTestStat}{{8.13}}
\newcommand{\CSnDefacementsCASignificanceTestPValue}{{p < .05}}

\newcommand{\CSnDefacementsCAEffectSize}{{0.03}}

\newcommand{\CSnDefacementsCAPostHocPValueONETWO}{{p = .0944}}

\newcommand{\CSnDefacementsCAPostHocPValueONETHREE}{{p = .2458}}

\newcommand{\CSnDefacementsCAPostHocPValueTWOTHREE}{{p < .01}}

\newcommand{\CSnDefacementsTRSignificanceTestDoF}{{2}}
\newcommand{\CSnDefacementsTRSignificanceTestStat}{{20.07}}
\newcommand{\CSnDefacementsTRSignificanceTestPValue}{{p < .0001}}

\newcommand{\CSnDefacementsTREffectSize}{{0.10}}

\newcommand{\CSnDefacementsTRPostHocPValueONETWO}{{p = .1171}}

\newcommand{\CSnDefacementsTRPostHocPValueONETHREE}{{p < .0001}}

\newcommand{\CSnDefacementsTRPostHocPValueTWOTHREE}{{p < .05}}

\newcommand{\CSnDefacementsSGSignificanceTestDoF}{{2}}
\newcommand{\CSnDefacementsSGSignificanceTestStat}{{3.83}}
\newcommand{\CSnDefacementsSGSignificanceTestPValue}{{p = .1473}}

\newcommand{\CSnDefacementsSGEffectSize}{{0.01}}

\newcommand{\CSnDefacementsSGPostHocPValueONETWO}{{p = .7583}}

\newcommand{\CSnDefacementsSGPostHocPValueONETHREE}{{p = .0677}}

\newcommand{\CSnDefacementsSGPostHocPValueTWOTHREE}{{p = .2085}}

\newcommand{\CSnDefacersRUSignificanceTestDoF}{{2}}
\newcommand{\CSnDefacersRUSignificanceTestStat}{{26.57}}
\newcommand{\CSnDefacersRUSignificanceTestPValue}{{p < .0001}}

\newcommand{\CSnDefacersRUEffectSize}{{0.14}}

\newcommand{\CSnDefacersRUPostHocPValueONETWO}{{p < .0001}}

\newcommand{\CSnDefacersRUPostHocPValueONETHREE}{{p = .9083}}

\newcommand{\CSnDefacersRUPostHocPValueTWOTHREE}{{p < .0001}}

\newcommand{\CSnDefacersUASignificanceTestDoF}{{2}}
\newcommand{\CSnDefacersUASignificanceTestStat}{{13.64}}
\newcommand{\CSnDefacersUASignificanceTestPValue}{{p < .01}}

\newcommand{\CSnDefacersUAEffectSize}{{0.07}}

\newcommand{\CSnDefacersUAPostHocPValueONETWO}{{p < .01}}

\newcommand{\CSnDefacersUAPostHocPValueONETHREE}{{p = .9286}}

\newcommand{\CSnDefacersUAPostHocPValueTWOTHREE}{{p < .001}}

\newcommand{\CSnDefacersUSSignificanceTestDoF}{{2}}
\newcommand{\CSnDefacersUSSignificanceTestStat}{{24.30}}
\newcommand{\CSnDefacersUSSignificanceTestPValue}{{p < .0001}}

\newcommand{\CSnDefacersUSEffectSize}{{0.13}}

\newcommand{\CSnDefacersUSPostHocPValueONETWO}{{p < .001}}

\newcommand{\CSnDefacersUSPostHocPValueONETHREE}{{p = .5961}}

\newcommand{\CSnDefacersUSPostHocPValueTWOTHREE}{{p < .0001}}

\newcommand{\CSnDefacersBRSignificanceTestDoF}{{2}}
\newcommand{\CSnDefacersBRSignificanceTestStat}{{11.68}}
\newcommand{\CSnDefacersBRSignificanceTestPValue}{{p < .01}}

\newcommand{\CSnDefacersBREffectSize}{{0.05}}

\newcommand{\CSnDefacersBRPostHocPValueONETWO}{{p = .3405}}

\newcommand{\CSnDefacersBRPostHocPValueONETHREE}{{p < .05}}

\newcommand{\CSnDefacersBRPostHocPValueTWOTHREE}{{p < .01}}

\newcommand{\CSnDefacersDESignificanceTestDoFBetween}{{2}}
\newcommand{\CSnDefacersDESignificanceTestDoFWithin}{{178}}
\newcommand{\CSnDefacersDESignificanceTestStat}{{3.24}}
\newcommand{\CSnDefacersDESignificanceTestPValue}{{p < .05}}

\newcommand{\CSnDefacersDEEffectSize}{{0.04}}

\newcommand{\CSnDefacersDEPostHocPValueONETWO}{{p = .7584}}

\newcommand{\CSnDefacersDEPostHocPValueONETHREE}{{p = .1858}}

\newcommand{\CSnDefacersDEPostHocPValueTWOTHREE}{{p = .0568}}

\newcommand{\CSnDefacersINSignificanceTestDoF}{{2}}
\newcommand{\CSnDefacersINSignificanceTestStat}{{3.90}}
\newcommand{\CSnDefacersINSignificanceTestPValue}{{p = .1424}}

\newcommand{\CSnDefacersINEffectSize}{{0.01}}

\newcommand{\CSnDefacersINPostHocPValueONETWO}{{p = .0746}}

\newcommand{\CSnDefacersINPostHocPValueONETHREE}{{p = .8734}}

\newcommand{\CSnDefacersINPostHocPValueTWOTHREE}{{p = .0704}}

\newcommand{\CSnDefacersIDSignificanceTestDoF}{{2}}
\newcommand{\CSnDefacersIDSignificanceTestStat}{{17.93}}
\newcommand{\CSnDefacersIDSignificanceTestPValue}{{p < .001}}

\newcommand{\CSnDefacersIDEffectSize}{{0.09}}

\newcommand{\CSnDefacersIDPostHocPValueONETWO}{{p < .001}}

\newcommand{\CSnDefacersIDPostHocPValueONETHREE}{{p = .5517}}

\newcommand{\CSnDefacersIDPostHocPValueTWOTHREE}{{p < .001}}

\newcommand{\CSnDefacersCASignificanceTestDoF}{{2}}
\newcommand{\CSnDefacersCASignificanceTestStat}{{9.51}}
\newcommand{\CSnDefacersCASignificanceTestPValue}{{p < .01}}

\newcommand{\CSnDefacersCAEffectSize}{{0.04}}

\newcommand{\CSnDefacersCAPostHocPValueONETWO}{{p = .1020}}

\newcommand{\CSnDefacersCAPostHocPValueONETHREE}{{p = .1492}}

\newcommand{\CSnDefacersCAPostHocPValueTWOTHREE}{{p < .01}}

\newcommand{\CSnDefacersTRSignificanceTestDoF}{{2}}
\newcommand{\CSnDefacersTRSignificanceTestStat}{{13.26}}
\newcommand{\CSnDefacersTRSignificanceTestPValue}{{p < .01}}

\newcommand{\CSnDefacersTREffectSize}{{0.06}}

\newcommand{\CSnDefacersTRPostHocPValueONETWO}{{p = .5501}}

\newcommand{\CSnDefacersTRPostHocPValueONETHREE}{{p < .001}}

\newcommand{\CSnDefacersTRPostHocPValueTWOTHREE}{{p < .05}}

\newcommand{\CSnDefacersSGSignificanceTestDoF}{{2}}
\newcommand{\CSnDefacersSGSignificanceTestStat}{{5.90}}
\newcommand{\CSnDefacersSGSignificanceTestPValue}{{p = .0524}}

\newcommand{\CSnDefacersSGEffectSize}{{0.02}}

\newcommand{\CSnDefacersSGPostHocPValueONETWO}{{p = .3056}}

\newcommand{\CSnDefacersSGPostHocPValueONETHREE}{{p < .05}}

\newcommand{\CSnDefacersSGPostHocPValueTWOTHREE}{{p = .3218}}

\newcommand{\CSnDDoSVictimsRUSignificanceTestDoF}{{2}}
\newcommand{\CSnDDoSVictimsRUSignificanceTestStat}{{57.13}}
\newcommand{\CSnDDoSVictimsRUSignificanceTestPValue}{{p < .0001}}

\newcommand{\CSnDDoSVictimsRUEffectSize}{{0.31}}

\newcommand{\CSnDDoSVictimsRUPostHocPValueONETWO}{{p < .0001}}

\newcommand{\CSnDDoSVictimsRUPostHocPValueONETHREE}{{p < .01}}

\newcommand{\CSnDDoSVictimsRUPostHocPValueTWOTHREE}{{p < .0001}}

\newcommand{\CSnDDoSVictimsUASignificanceTestDoF}{{2}}
\newcommand{\CSnDDoSVictimsUASignificanceTestStat}{{15.16}}
\newcommand{\CSnDDoSVictimsUASignificanceTestPValue}{{p < .001}}

\newcommand{\CSnDDoSVictimsUAEffectSize}{{0.07}}

\newcommand{\CSnDDoSVictimsUAPostHocPValueONETWO}{{p < .001}}

\newcommand{\CSnDDoSVictimsUAPostHocPValueONETHREE}{{p = .8765}}

\newcommand{\CSnDDoSVictimsUAPostHocPValueTWOTHREE}{{p < .001}}

\newcommand{\CSnDDoSVictimsUSSignificanceTestDoF}{{2}}
\newcommand{\CSnDDoSVictimsUSSignificanceTestStat}{{4.43}}
\newcommand{\CSnDDoSVictimsUSSignificanceTestPValue}{{p = .1093}}

\newcommand{\CSnDDoSVictimsUSEffectSize}{{0.01}}

\newcommand{\CSnDDoSVictimsUSPostHocPValueONETWO}{{p = .2527}}

\newcommand{\CSnDDoSVictimsUSPostHocPValueONETHREE}{{p < .05}}

\newcommand{\CSnDDoSVictimsUSPostHocPValueTWOTHREE}{{p = .5594}}

\newcommand{\CSnDDoSVictimsBRSignificanceTestDoF}{{2}}
\newcommand{\CSnDDoSVictimsBRSignificanceTestStat}{{13.12}}
\newcommand{\CSnDDoSVictimsBRSignificanceTestPValue}{{p < .01}}

\newcommand{\CSnDDoSVictimsBREffectSize}{{0.06}}

\newcommand{\CSnDDoSVictimsBRPostHocPValueONETWO}{{p < .001}}

\newcommand{\CSnDDoSVictimsBRPostHocPValueONETHREE}{{p < .05}}

\newcommand{\CSnDDoSVictimsBRPostHocPValueTWOTHREE}{{p < .05}}

\newcommand{\CSnDDoSVictimsDESignificanceTestDoF}{{2}}
\newcommand{\CSnDDoSVictimsDESignificanceTestStat}{{17.24}}
\newcommand{\CSnDDoSVictimsDESignificanceTestPValue}{{p < .001}}

\newcommand{\CSnDDoSVictimsDEEffectSize}{{0.09}}

\newcommand{\CSnDDoSVictimsDEPostHocPValueONETWO}{{p < .001}}

\newcommand{\CSnDDoSVictimsDEPostHocPValueONETHREE}{{p < .001}}

\newcommand{\CSnDDoSVictimsDEPostHocPValueTWOTHREE}{{p = .2039}}

\newcommand{\CSnDDoSVictimsBDSignificanceTestDoF}{{2}}
\newcommand{\CSnDDoSVictimsBDSignificanceTestStat}{{4.43}}
\newcommand{\CSnDDoSVictimsBDSignificanceTestPValue}{{p = .1090}}

\newcommand{\CSnDDoSVictimsBDEffectSize}{{0.01}}

\newcommand{\CSnDDoSVictimsBDPostHocPValueONETWO}{{p < .05}}

\newcommand{\CSnDDoSVictimsBDPostHocPValueONETHREE}{{p = .1353}}

\newcommand{\CSnDDoSVictimsBDPostHocPValueTWOTHREE}{{p = .3585}}

\newcommand{\CSnDDoSVictimsCNSignificanceTestDoF}{{2}}
\newcommand{\CSnDDoSVictimsCNSignificanceTestStat}{{65.91}}
\newcommand{\CSnDDoSVictimsCNSignificanceTestPValue}{{p < .0001}}

\newcommand{\CSnDDoSVictimsCNEffectSize}{{0.36}}

\newcommand{\CSnDDoSVictimsCNPostHocPValueONETWO}{{p < .0001}}

\newcommand{\CSnDDoSVictimsCNPostHocPValueONETHREE}{{p < .0001}}

\newcommand{\CSnDDoSVictimsCNPostHocPValueTWOTHREE}{{p = .0674}}

\newcommand{\CSnDDoSVictimsFRSignificanceTestDoF}{{2}}
\newcommand{\CSnDDoSVictimsFRSignificanceTestStat}{{9.96}}
\newcommand{\CSnDDoSVictimsFRSignificanceTestPValue}{{p < .01}}

\newcommand{\CSnDDoSVictimsFREffectSize}{{0.04}}

\newcommand{\CSnDDoSVictimsFRPostHocPValueONETWO}{{p = .1519}}

\newcommand{\CSnDDoSVictimsFRPostHocPValueONETHREE}{{p < .01}}

\newcommand{\CSnDDoSVictimsFRPostHocPValueTWOTHREE}{{p = .2366}}

\newcommand{\CSnDDoSVictimsGBSignificanceTestDoF}{{2}}
\newcommand{\CSnDDoSVictimsGBSignificanceTestStat}{{7.94}}
\newcommand{\CSnDDoSVictimsGBSignificanceTestPValue}{{p < .05}}

\newcommand{\CSnDDoSVictimsGBEffectSize}{{0.03}}

\newcommand{\CSnDDoSVictimsGBPostHocPValueONETWO}{{p = .5258}}

\newcommand{\CSnDDoSVictimsGBPostHocPValueONETHREE}{{p < .01}}

\newcommand{\CSnDDoSVictimsGBPostHocPValueTWOTHREE}{{p = .0976}}

\newcommand{\CSnDDoSVictimsPLSignificanceTestDoF}{{2}}
\newcommand{\CSnDDoSVictimsPLSignificanceTestStat}{{0.04}}
\newcommand{\CSnDDoSVictimsPLSignificanceTestPValue}{{p = .9809}}

\newcommand{\CSnDDoSVictimsPLEffectSize}{{0.00}}

\newcommand{\CSnDDoSVictimsPLPostHocPValueONETWO}{{p = .9081}}

\newcommand{\CSnDDoSVictimsPLPostHocPValueONETHREE}{{p = .9365}}

\newcommand{\CSnDDoSVictimsPLPostHocPValueTWOTHREE}{{p = .8449}}

\newcommand{\CSnDDoSAttacksRUSignificanceTestDoF}{{2}}
\newcommand{\CSnDDoSAttacksRUSignificanceTestStat}{{60.67}}
\newcommand{\CSnDDoSAttacksRUSignificanceTestPValue}{{p < .0001}}

\newcommand{\CSnDDoSAttacksRUEffectSize}{{0.33}}

\newcommand{\CSnDDoSAttacksRUPostHocPValueONETWO}{{p < .0001}}

\newcommand{\CSnDDoSAttacksRUPostHocPValueONETHREE}{{p < .0001}}

\newcommand{\CSnDDoSAttacksRUPostHocPValueTWOTHREE}{{p < .0001}}

\newcommand{\CSnDDoSAttacksUASignificanceTestDoF}{{2}}
\newcommand{\CSnDDoSAttacksUASignificanceTestStat}{{12.59}}
\newcommand{\CSnDDoSAttacksUASignificanceTestPValue}{{p < .01}}

\newcommand{\CSnDDoSAttacksUAEffectSize}{{0.06}}

\newcommand{\CSnDDoSAttacksUAPostHocPValueONETWO}{{p < .01}}

\newcommand{\CSnDDoSAttacksUAPostHocPValueONETHREE}{{p = .4593}}

\newcommand{\CSnDDoSAttacksUAPostHocPValueTWOTHREE}{{p < .001}}

\newcommand{\CSnDDoSAttacksUSSignificanceTestDoF}{{2}}
\newcommand{\CSnDDoSAttacksUSSignificanceTestStat}{{6.98}}
\newcommand{\CSnDDoSAttacksUSSignificanceTestPValue}{{p < .05}}

\newcommand{\CSnDDoSAttacksUSEffectSize}{{0.03}}

\newcommand{\CSnDDoSAttacksUSPostHocPValueONETWO}{{p = .5592}}

\newcommand{\CSnDDoSAttacksUSPostHocPValueONETHREE}{{p < .05}}

\newcommand{\CSnDDoSAttacksUSPostHocPValueTWOTHREE}{{p = .1182}}

\newcommand{\CSnDDoSAttacksBRSignificanceTestDoF}{{2}}
\newcommand{\CSnDDoSAttacksBRSignificanceTestStat}{{9.81}}
\newcommand{\CSnDDoSAttacksBRSignificanceTestPValue}{{p < .01}}

\newcommand{\CSnDDoSAttacksBREffectSize}{{0.04}}

\newcommand{\CSnDDoSAttacksBRPostHocPValueONETWO}{{p < .01}}

\newcommand{\CSnDDoSAttacksBRPostHocPValueONETHREE}{{p = .2006}}

\newcommand{\CSnDDoSAttacksBRPostHocPValueTWOTHREE}{{p < .05}}

\newcommand{\CSnDDoSAttacksDESignificanceTestDoF}{{2}}
\newcommand{\CSnDDoSAttacksDESignificanceTestStat}{{9.49}}
\newcommand{\CSnDDoSAttacksDESignificanceTestPValue}{{p < .01}}

\newcommand{\CSnDDoSAttacksDEEffectSize}{{0.04}}

\newcommand{\CSnDDoSAttacksDEPostHocPValueONETWO}{{p < .01}}

\newcommand{\CSnDDoSAttacksDEPostHocPValueONETHREE}{{p < .01}}

\newcommand{\CSnDDoSAttacksDEPostHocPValueTWOTHREE}{{p = .6609}}

\newcommand{\CSnDDoSAttacksBDSignificanceTestDoF}{{2}}
\newcommand{\CSnDDoSAttacksBDSignificanceTestStat}{{3.96}}
\newcommand{\CSnDDoSAttacksBDSignificanceTestPValue}{{p = .1379}}

\newcommand{\CSnDDoSAttacksBDEffectSize}{{0.01}}

\newcommand{\CSnDDoSAttacksBDPostHocPValueONETWO}{{p < .05}}

\newcommand{\CSnDDoSAttacksBDPostHocPValueONETHREE}{{p = .2198}}

\newcommand{\CSnDDoSAttacksBDPostHocPValueTWOTHREE}{{p = .2785}}

\newcommand{\CSnDDoSAttacksCNSignificanceTestDoF}{{2}}
\newcommand{\CSnDDoSAttacksCNSignificanceTestStat}{{80.16}}
\newcommand{\CSnDDoSAttacksCNSignificanceTestPValue}{{p < .0001}}

\newcommand{\CSnDDoSAttacksCNEffectSize}{{0.44}}

\newcommand{\CSnDDoSAttacksCNPostHocPValueONETWO}{{p < .001}}

\newcommand{\CSnDDoSAttacksCNPostHocPValueONETHREE}{{p < .0001}}

\newcommand{\CSnDDoSAttacksCNPostHocPValueTWOTHREE}{{p < .0001}}

\newcommand{\CSnDDoSAttacksFRSignificanceTestDoF}{{2}}
\newcommand{\CSnDDoSAttacksFRSignificanceTestStat}{{16.04}}
\newcommand{\CSnDDoSAttacksFRSignificanceTestPValue}{{p < .001}}

\newcommand{\CSnDDoSAttacksFREffectSize}{{0.08}}

\newcommand{\CSnDDoSAttacksFRPostHocPValueONETWO}{{p = .2586}}

\newcommand{\CSnDDoSAttacksFRPostHocPValueONETHREE}{{p < .001}}

\newcommand{\CSnDDoSAttacksFRPostHocPValueTWOTHREE}{{p < .05}}

\newcommand{\CSnDDoSAttacksGBSignificanceTestDoF}{{2}}
\newcommand{\CSnDDoSAttacksGBSignificanceTestStat}{{13.90}}
\newcommand{\CSnDDoSAttacksGBSignificanceTestPValue}{{p < .01}}

\newcommand{\CSnDDoSAttacksGBEffectSize}{{0.07}}

\newcommand{\CSnDDoSAttacksGBPostHocPValueONETWO}{{p = .4892}}

\newcommand{\CSnDDoSAttacksGBPostHocPValueONETHREE}{{p < .001}}

\newcommand{\CSnDDoSAttacksGBPostHocPValueTWOTHREE}{{p < .05}}

\newcommand{\CSnDDoSAttacksPLSignificanceTestDoF}{{2}}
\newcommand{\CSnDDoSAttacksPLSignificanceTestStat}{{0.34}}
\newcommand{\CSnDDoSAttacksPLSignificanceTestPValue}{{p = .8423}}

\newcommand{\CSnDDoSAttacksPLEffectSize}{{0.00}}

\newcommand{\CSnDDoSAttacksPLPostHocPValueONETWO}{{p = .8841}}

\newcommand{\CSnDDoSAttacksPLPostHocPValueONETHREE}{{p = .6759}}

\newcommand{\CSnDDoSAttacksPLPostHocPValueTWOTHREE}{{p = .6002}}

\newcommand{\CStopCategoriesProportion}{{80.21}}
\newcommand{\CSnTelegramPromotedTargets}{{3\,845}}
\newcommand{\CSnTelegramPromotedTargetsDomains}{{1\,554}}
\newcommand{\CSnTelegramPromotedTargetsDomainsProportion}{{40.42}}
\newcommand{\CSnTelegramPromotedTargetsIPs}{{2\,291}}
\newcommand{\CSnTelegramPromotedTargetsIPsProportion}{{59.58}}
\newcommand{\CSnTelegramPromotedTargetsOverlapsWithDDoS}{{707}}
\newcommand{\CSnTelegramPromotedTargetsOverlapsWithDDoSProportion}{{30.86}}
\newcommand{\CSnTelegramPromotedTargetsOverlapsWithDefacements}{{59}}
\newcommand{\CSnTelegramPromotedTargetsOverlapsWithDefacementsDomainOnly}{{7}}
\newcommand{\CSnTelegramPromotedTargetsOverlapsWithDefacementsDomainOnlyProportion}{{0.18}}
\newcommand{\CSnTelegramPromotedTargetsOverlapsWithDefacementsIPOnly}{{52}}
\newcommand{\CSnTelegramPromotedTargetsOverlapsWithDefacementsIPOnlyProportion}{{1.35}}
\newcommand{\CSnTelegramPromotedTargetsOverlapsWithDefacementsProportion}{{1.53}}

\newcommand{\CSnDefacementUnifiedValidRU}{{4\,340}}
\newcommand{\CSnDefacementUnifiedValidUA}{{1\,559}}
\newcommand{\CSnDefacementUnifiedValidRUUA}{{5\,899}}
\newcommand{\CSnDefacementUnifiedValidRUUAProportion}{{2.15}}
\newcommand{\CSnDefacementUnifiedValidRUUAGroupedByMessage}{{1\,341}}
\newcommand{\CSnDefacementWarRelatedFunMotives}{{2\,723}}
\newcommand{\CSnDefacementWarRelatedFunMotivesProps}{{46.16}}
\newcommand{\CSnDefacementWarRelatedSelfExpressionMotives}{{1\,219}}
\newcommand{\CSnDefacementWarRelatedSelfExpressionMotivesProps}{{20.66}}
\newcommand{\CSnDefacementWarRelatedPatriotiMotives}{{58}}
\newcommand{\CSnDefacementWarRelatedPatriotiMotivesProps}{{0.98}}
\newcommand{\CSnDefacementWarRelatedNationalisticConflictsMotives}{{143}}
\newcommand{\CSnDefacementWarRelatedNationalisticConflictsMotivesProps}{{2.42}}
\newcommand{\CSnDefacementWarRelatedProUAMotives}{{286}}
\newcommand{\CSnDefacementWarRelatedProUAMotivesProps}{{4.85}}
\newcommand{\CSnDefacementWarRelatedProUAMotivesButTargetUA}{{12}}

\newcommand{\CSnDefacementWarRelatedProRUMotives}{{103}}
\newcommand{\CSnDefacementWarRelatedProRUMotivesProps}{{1.75}}
\newcommand{\CSnDefacementWarRelatedProRUMotivesButTargetRU}{{22}}

\newcommand{\CSnDefacementWarRelatedFinanceMotives}{{89}}
\newcommand{\CSnDefacementWarRelatedFinanceMotivesProps}{{1.51}}
\newcommand{\CSnDefacementWarRelatedNoneDeterministicMotives}{{1\,278}}
\newcommand{\CSnDefacementWarRelatedNoneDeterministicMotivesProps}{{21.66}}
\newcommand{\CSnHFRUThreads}{{115}}
\newcommand{\CSnHFUAThreads}{{108}}
\newcommand{\CSnHFRUUAOverlapsThreads}{{100}}
\newcommand{\CSnHFRUUAFinalThreads}{{123}}
\newcommand{\CSnRUUAHighlyRelevantThreads}{{84}}
\newcommand{\CSnRUUAHighlyRelevantPosts}{{1\,279}}
\newcommand{\CSnRUUALowerRelevantThreads}{{39}}
\newcommand{\CSnRUUALowerRelevantPosts}{{485}}
\newcommand{\CSnRUUAAllRelevantPosts}{{1\,764}}
\newcommand{\CSnRUUAAllRelevantUsers}{{372}}

\newcommand{\CSnRUUARelatedPostsSignificanceTestMethod}{{Kruskal-Wallis}}
\newcommand{\CSnRUUARelatedPostsSignificanceTestDoF}{{2}}
\newcommand{\CSnRUUARelatedPostsSignificanceTestStat}{{72.98}}
\newcommand{\CSnRUUARelatedPostsSignificanceTestPValue}{{p < .0001}}

\newcommand{\CSnRUUARelatedPostsEffectSize}{{0.40}}

\newcommand{\CSnRUUARelatedPostsPostHocPValueONETHREE}{{p = .8501}}

\newcommand{\CSnRUUARelatedPostsPostHocPValueTWOTHREE}{{p < .0001}}

\newcommand{\CSnRUUARelatedUsersSignificanceTestMethod}{{Kruskal-Wallis}}
\newcommand{\CSnRUUARelatedUsersSignificanceTestDoF}{{2}}
\newcommand{\CSnRUUARelatedUsersSignificanceTestStat}{{77.54}}
\newcommand{\CSnRUUARelatedUsersSignificanceTestPValue}{{p < .0001}}

\newcommand{\CSnRUUARelatedUsersEffectSize}{{0.42}}

\newcommand{\CSnRUUARelatedUsersPostHocPValueONETHREE}{{p = .6657}}

\newcommand{\CSnRUUARelatedUsersPostHocPValueTWOTHREE}{{p < .0001}}

\newcommand{\CSnHFRUUANPostsTopCategoriesProportion}{{97.22}}

\newcommand{\CSnHFRUUANPostsTopCategoriesNumberOneProportion}{{53.40}}

\newcommand{\CSnHFRUUANPostsTopCategoriesNumberTwoProportion}{{33.28}}

\usepackage{contour,ulem}

\contourlength{0.8pt}

\usepackage{tikz}

\newcommand{\tikzcmark}{%
\tikz[scale=0.23] {
    \draw[line width=0.7,line cap=round] (0.25,0) to [bend left=10] (1,1);
    \draw[line width=0.8,line cap=round] (0,0.35) to [bend right=1] (0.23,0);
}}
\newcommand{\dashcmark}{
    \textcolor{black!20}{\tikzcmark}
}

\usepackage{algorithm,algpseudocode,algorithmicx}
\algnewcommand\algorithmicforeach{\textbf{for each}}
\algdef{S}[FOR]{ForEach}[1]{\algorithmicforeach\ #1\ \algorithmicdo}
\algdef{SE}[DOWHILE]{Do}{doWhile}{\algorithmicdo}[1]{\algorithmicwhile\ #1}

\usepackage{booktabs} % For professional looking tables
\usepackage{multirow,multicol} % for multiple columns/rows
\usepackage{makecell} % making multiple lines in cells
\usepackage[switch]{lineno} % for line numbers
\usepackage{threeparttable} % to have note in tables
\usepackage{url}  % break long URLs
\usepackage{xspace} % for user defined commands
\usepackage{colortbl} % to modify colours of columns
\usepackage{fancyhdr} % to add header and footer
\DeclareUrlCommand\myurl{\urlstyle{tt}} % set URLs with different fonts
\newcommand{\para}[1]{\vspace{1mm}\noindent\textbf{#1.}}
\usepackage{adjustbox} % for rotating table fields
\newcolumntype{R}[2]{>{\adjustbox{angle=#1,lap=\width-(#2)}\bgroup}l<{\egroup}}

\newcommand{\hackforums}{{\small\scshape Hack Forums}\xspace}

\newcommand{\zoneh}{{\small\scshape Zone-H}\xspace}

\newcommand{\zonexsec}{{\small\scshape Zone-Xsec}\xspace}

\newcommand{\haxorid}{{\small\scshape Haxor-ID}\xspace}

\newcommand{\defacerpro}{{\small\scshape Defacer-Pro}\xspace}

\newcommand{\defacerid}{{\small\scshape Defacer-ID}\xspace}
\newcommand{\defaceridsmall}{{\footnotesize\scshape Defacer-ID}\xspace}
\newcommand{\ownzyou}{{\small\scshape OwnzYou}\xspace}

\newcommand{\hackmirror}{{\small\scshape Hack Mirror}\xspace}

\newcommand{\mirrorzone}{{\small\scshape Mirror Zone}\xspace}

\newcommand{\hackcn}{{\small\scshape Hack-CN}\xspace}

\newcommand{\mydeface}{{\small\scshape MyDeface}\xspace}

\newcommand{\crimebb}{{\small\scshape CrimeBB}\xspace}

\newcommand{\ccc}{Cambridge Cybercrime Centre\xspace}

\newcommand{\captcha}{{\small\scshape Captcha}\xspace}

\newcommand{\ddosguard}{DDoS-Guard\xspace}

\newcommand{\hsAttackDurationTwentyTwentyTwoFiftyP}{1.53}

\newcommand{\hsAttackDurationTwentyTwentyTwoOneHundredP}{11\,300}

\newcommand{\hsObsAttacksWeekTwentyTwentyTwoCINF}{95\% CI [11\,900, 271\,000]}

\newcommand{\hsObsAttacksWeekTwentyTwentyTwoMedian}{35\,000}

\newcommand{\hsPrefixAttacksPerWeekTwentyTwentyTwoCINF}{95\% CI [0, 3\,480]}

\newcommand{\hsPrefixAttacksPerWeekTwentyTwentyTwoMedian}{438}

\newcommand{\hsSensorsDayTwentyTwentyTwoCINF}{95\% CI [34, 51]}

\newcommand{\hsSensorsDayTwentyTwentyTwoMedian}{50}

\AtBeginDocument{ % typeset BibTeX logo in the docs
  \providecommand\BibTeX{{%
    \normalfont B\kern-0.5em{\scshape i\kern-0.25em b}\kern-0.8em\TeX}}}

\copyrightyear{2024}
\acmYear{2024}
\setcopyright{rightsretained}
\acmConference[WWW '24]{Proceedings of the ACM Web Conference 2024}{May 13--17, 2024}{Singapore, Singapore}
\acmBooktitle{Proceedings of the ACM Web Conference 2024 (WWW '24), May 13--17, 2024, Singapore, Singapore}
\acmDOI{10.1145/3589334.3645401}
\acmISBN{979-8-4007-0171-9/24/05}
\makeatletter
\gdef\@copyrightpermission{
  \begin{minipage}{0.3\columnwidth}
   \href{https://creativecommons.org/licenses/by/4.0/}{\includegraphics[width=0.90\textwidth]{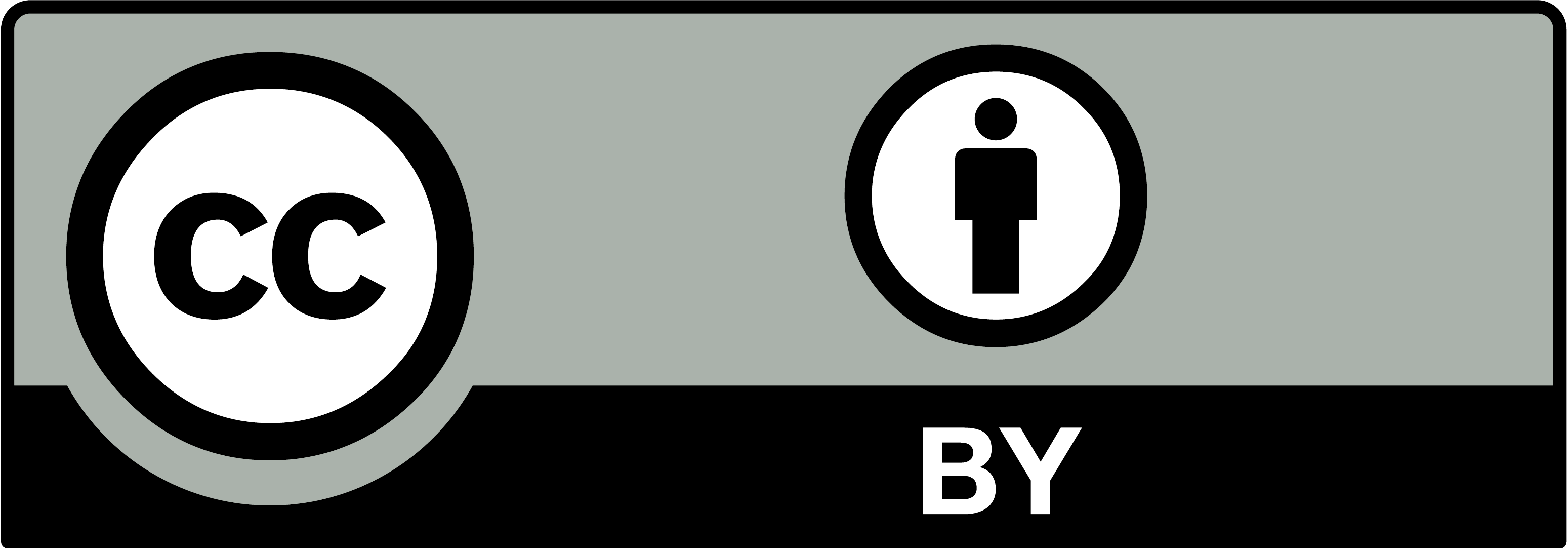}}
  \end{minipage}\hfill
  \begin{minipage}{0.7\columnwidth}
   \href{https://creativecommons.org/licenses/by/4.0/}{This work is licensed under a Creative Commons Attribution International 4.0 License.}
  \end{minipage}
  \vspace{5pt}
}
\makeatother

\begin{document}
\title[Getting Bored of Cyberwar: Exploring the Role of Low-level Cybercrime Actors in the Russia-Ukraine Conflict]{Getting Bored of Cyberwar: Exploring the Role of Low-level Cybercrime Actors in the Russia-Ukraine Conflict}

\author{Anh V. Vu} \orcid{0000-0002-6287-0226}
\affiliation{\institution{University of Cambridge} \city{Cambridge} \country{United Kingdom}}
\email{anh.vu@cl.cam.ac.uk}
\author{Daniel R. Thomas} \orcid{0000-0001-8936-0683}
\affiliation{\institution{University of Strathclyde} \city{Glasgow} \country{United Kingdom}}
\email{d.thomas@strath.ac.uk}
\author{Ben Collier} \orcid{0000-0002-9207-3068}
\affiliation{\institution{University of Edinburgh} \city{Edinburgh} \country{United Kingdom}}
\email{ben.collier@ed.ac.uk}
\author{Alice Hutchings} \orcid{0000-0003-3037-2684}
\affiliation{\institution{University of Cambridge} \city{Cambridge} \country{United Kingdom}}
\email{alice.hutchings@cl.cam.ac.uk}
\author{Richard Clayton} \orcid{0000-0002-1673-918X}
\affiliation{\institution{University of Cambridge} \city{Cambridge} \country{United Kingdom}}
\email{richard.clayton@cl.cam.ac.uk}
\author{Ross Anderson} \orcid{0000-0001-8697-5682}
\affiliation{\institution{University of Cambridge \& Edinburgh} \city{Cambridge} \country{United Kingdom}}
\email{ross.anderson@cl.cam.ac.uk}
\renewcommand{\shortauthors}{Anh V. Vu et al.}

\begin{abstract}
There has been substantial commentary on the role of cyberattacks carried out by low-level cybercrime actors in the Russia-Ukraine conflict. We analyse \CSnDefacementRawTotalRounded~website defacement attacks, \CSnDDoSAttacksRounded~UDP amplification DDoS attacks, \CSnRUUAAllRelevantPosts~posts made by \CSnRUUAAllRelevantUsers~users on \hackforums mentioning the two countries, and \CSnTelegramMessages~Telegram announcements (with \CSnTelegramRepliesRounded~replies) of a volunteer hacking group for two months before and four months after the invasion. We find the conflict briefly but notably caught the attention of low-level cybercrime actors, with significant increases in online discussion and both types of attacks targeting Russia and Ukraine. However, there was little evidence of high-profile actions; the role of these players in the ongoing hybrid warfare is minor, and they should be separated from persistent and motivated `hacktivists' in state-sponsored operations. Their involvement in the conflict appears to have been short-lived and fleeting, with a clear loss of interest in discussing the situation and carrying out both website defacement and DDoS attacks against either Russia or Ukraine after just a few weeks.
\end{abstract}

\begin{CCSXML}
<ccs2012>
   <concept>
       <concept_id>10003456.10003462.10003574</concept_id>
       <concept_desc>Social and professional topics~Computer crime</concept_desc>
       <concept_significance>500</concept_significance>
       </concept>
   <concept>
       <concept_id>10010405.10010476.10010478.10003600</concept_id>
       <concept_desc>Applied computing~Cyberwarfare</concept_desc>
       <concept_significance>500</concept_significance>
       </concept>
   <concept>
       <concept_id>10002978.10003029.10003032</concept_id>
       <concept_desc>Security and privacy~Social aspects of security and privacy</concept_desc>
       <concept_significance>500</concept_significance>
       </concept>
   <concept>
       <concept_id>10003033.10003083.10003014.10011610</concept_id>
       <concept_desc>Networks~Denial-of-service attacks</concept_desc>
       <concept_significance>500</concept_significance>
       </concept>
   <concept>
       <concept_id>10002950.10003648.10003688.10003693</concept_id>
       <concept_desc>Mathematics of computing~Time series analysis</concept_desc>
       <concept_significance>100</concept_significance>
       </concept>
 </ccs2012>
\end{CCSXML}
\ccsdesc[500]{Social and professional topics~Computer crime}
\ccsdesc[500]{Applied computing~Cyberwarfare}
\ccsdesc[100]{Security and privacy~Social aspects of security and privacy}
\ccsdesc[100]{Networks~Denial-of-service attacks}
\ccsdesc[100]{Mathematics of computing~Time series analysis}

\keywords{Russia-Ukraine conflict; DDoS attacks; website defacement attacks; cybercrime; cyberwar; volunteer hacktivists; IT Army of Ukraine.}

\maketitle

\section{Introduction} \label{sec:introduction}
Researchers, politicians, and journalists have long been fascinated by `cyberwar' -- the spectre of armed conflict between nations spilling over into attacks conducted over the Internet~\cite{rid2012cyber}. `Colder' forms of inter-state conflict are characterised by espionage and intelligence gathering, which may facilitate the degradation of online systems once hostilities commence~\cite{everett2013optically}. Alongside this, there has been a thirty-year history of speculation around how the tools and techniques of the cybercrime underground -- Distributed Denial of Service (DDoS) attacks, disruption and compromise of services, web defacements, and similar techniques -- might allow civilians to play a role in a `hot' war between developed nations~\cite{arquilla1993cyberwar}. Much of this speculation, drawing from criminological models of low-level cybercrime groups and on links between this underground and well-organised `hacktivist' movements, has argued these groups would play a crucial role, making the future of war hybrid, chaotic, and unpredictable~\cite{wall2007cybercrime}. The 2022 Russian invasion of Ukraine provides an opportunity to assess what has happened in practice.

Russia and Ukraine have a long history of electronic information warfare~\cite{jaitner2015russian} and are among the most active cybercrime hubs~\cite{lusthaus2020mapping}. When Russia invaded Ukraine on 24 February 2022, war-related attacks on the two countries were regularly reported~\cite{cyberattacknytimes}. A popular narrative is that the engagement of low-level cybercrime actors and volunteers could be a game changer and could undermine Russia's war~\cite{foreignpolicyitarmy}. Some commentators predicted it will be the first full-scale cyberwar~\cite{fullscalecyberwar}, its effects will last for decades~\cite{serpanos2022cyberwarfare}, and youngsters would be drawn into a `cyberwar' by joining IT Army of Ukraine -- a group backed by the Ukrainian state to co-ordinate volunteers and civilians to help disrupt Russian assets~\cite{soesantoitarmy,uagovernmentcreatingitarmy}. Some have suggested a `real cyberwar', predicting hacktivist attacks on Russia would escalate further through 2022~\cite{ddosattacks30fold}. These narratives regularly appear in the press and play a role in shaping domestic policy responses to cybercrime. Although less likely to grab headlines, a contrary narrative around `overhyped cyberwar' suggests cyber operations in the conflict have been slow~\cite{maschmeyer2022goodbye} and insignificant~\cite{kostyuk2022cyber}, while claims of an unprecedented level of cyberattacks and Russia's much-vaunted cyber capabilities are questionable~\cite{gibney2022russia,givens2023putin,willett2022cyber}. GCHQ commented the cyber conflict had not yet materialised~\cite{cyberwaroverhypedGCHQ} and pointed to the resilience of Ukraine's defences~\cite{russialoosinginformationwarfareGCHQ}.

Government-backed cyber operations~\cite{fogofwar,microsoftreport} and destructive attacks have continued~\cite{uareceiveddatawiping,microsoftdefending}. However, data about nation-state attacks is hard for academics to access, and actors behind significant real-world attacks tend to take steps to avoid scrutiny. We are particularly interested in non-governmental activity contributed by many low-level but high-volume actors, focusing on the hypothetical `volunteer army', where participants are mostly unskilled and their activity highly relies on off-the-shelf tools. We explore their role in the `cyberwar' between Russia and Ukraine, in which both sides have substantial IT infrastructure, a thriving digital underground crime ecosystem, and significant access to offensive capacities.

We longitudinally and statistically measure activities linked with low-level cybercrime actors, including web defacements (\S\ref{sec:defacements-evidence}) and DDoS attacks (\S\ref{sec:ddos-attacks-evidence}). The findings are incorporated with analyses of hacking community reactions, including a pro-Ukraine volunteer group~(\S\ref{sec:community-discussions}). The role of these actors in the conflict is discussed in~\S\ref{sec:discussion}. This study was ethically approved (see Appendix~\S\ref{appendix:ethical-issues}). All data and scripts are available to researchers on request (see Appendix~\S\ref{appendix:data-licensing}).

\begin{table*}[t]
\centering
\caption{The complete collection of the 5 most popular defacement archives for 6 months from 1 January 2022 to 30 June 2022.}
\setlength{\tabcolsep}{0.825em}
\small
\begin{tabular}{lrrrrrr}
\toprule
\textbf{} & \textbf{\zoneh} & \textbf{\ownzyou} & \textbf{\zonexsec} & \textbf{\haxorid} & \textbf{\defacerpro} & \textbf{Total} \\
Archive URL & zone-h.org & ownzyou.com & zone-xsec.com & hax.or.id & defacer.pro & 5 archives \\
\midrule
Manual staff verification & \tikzcmark & $\cdotp$ & $\cdotp$ & $\cdotp$ & $\cdotp$ & \dashcmark \\
Automatic validity sanitisation & $\cdotp$ & $\cdotp$ & \tikzcmark & \tikzcmark & \tikzcmark & \dashcmark \\
Team information & $\cdotp$ & $\cdotp$ & \tikzcmark & \tikzcmark & \tikzcmark & \dashcmark \\
Country of targeted victims & \tikzcmark & \tikzcmark & \tikzcmark & \tikzcmark & \tikzcmark & \tikzcmark\\
Originating country of defacers & $\cdotp$ & $\cdotp$ & $\cdotp$ & \tikzcmark & $\cdotp$ & \dashcmark\\
Reasons and/or motivations & $\cdotp$ & $\cdotp$ & $\cdotp$ & \tikzcmark & $\cdotp$ & \dashcmark\\
Types of vulnerability & $\cdotp$ & $\cdotp$ & $\cdotp$ & \tikzcmark & $\cdotp$ & \dashcmark\\
Snapshots of defaced websites & \tikzcmark & \tikzcmark & \tikzcmark & \tikzcmark & \tikzcmark & \tikzcmark\\
\arrayrulecolor{black!20}
\midrule
Defacements (raw) & \CSnDefacementRawZoneH & \CSnDefacementRawOwnzYou & \CSnDefacementRawZoneXSec & \CSnDefacementRawHaxorID & \CSnDefacementRawDefacerPro & \CSnDefacementRawTotal\\
Defacements~\textsuperscript{$\dagger$} & \CSnDefacementUnifiedZoneH & \CSnDefacementUnifiedOwnzYou & \CSnDefacementUnifiedZoneXSec & \CSnDefacementUnifiedHaxorID & \CSnDefacementUnifiedDefacerPro & \CSnDefacementUnifiedTotal\\
Valid defacements~\textsuperscript{$\dagger$} & \CSnDefacementUnifiedValidZoneH~(\CSnDefacementUnifiedValidProportionZoneH\%) & \CSnDefacementUnifiedValidOwnzYou~(\CSnDefacementUnifiedValidProportionOwnzYou\%) & \CSnDefacementUnifiedValidZoneXSec~(\CSnDefacementUnifiedValidProportionZoneXSec\%) & \CSnDefacementUnifiedValidHaxorID~(\CSnDefacementUnifiedValidProportionHaxorID\%) & \CSnDefacementUnifiedValidDefacerPro~(\CSnDefacementUnifiedValidProportionDefacerPro\%) & \CSnDefacementUnifiedValidTotal~(\CSnDefacementUnifiedValidProportionTotal\%)\\
Invalid defacement~\textsuperscript{$\dagger$} & \CSnDefacementUnifiedInvalidZoneH~(\CSnDefacementUnifiedInvalidProportionZoneH\%) & \CSnDefacementUnifiedInvalidOwnzYou~(\CSnDefacementUnifiedInvalidProportionOwnzYou\%) & \CSnDefacementUnifiedInvalidZoneXSec~(\CSnDefacementUnifiedInvalidProportionZoneXSec\%) & \CSnDefacementUnifiedInvalidHaxorID~(\CSnDefacementUnifiedInvalidProportionHaxorID\%) & \CSnDefacementUnifiedInvalidDefacerPro~(\CSnDefacementUnifiedInvalidProportionDefacerPro\%) & \CSnDefacementUnifiedInvalidTotal~(\CSnDefacementUnifiedInvalidProportionTotal\%)\\
\arrayrulecolor{black!20}
Defacers (raw) & \CSnDefacersRawZoneH & \CSnDefacersRawOwnzYou & \CSnDefacersRawZoneXSec & \CSnDefacersRawHaxorID & \CSnDefacersRawDefacerPro & \CSnDefacersRawTotal\\
Defacers~\textsuperscript{$\dagger$} & \CSnTotalHandlesZoneH & \CSnTotalHandlesOwnzYou & \CSnTotalHandlesZoneXSec & \CSnTotalHandlesHaxorID & \CSnTotalHandlesDefacerPro & \CSnTotalHandlesTotal\\
Defacers with valid reports~\textsuperscript{$\dagger$} & \CSnValidDefacersZoneH~(\CSnValidDefacersProportionZoneH\%) & \CSnValidDefacersOwnzYou~(\CSnValidDefacersProportionOwnzYou\%) & \CSnValidDefacersZoneXSec~(\CSnValidDefacersProportionZoneXSec\%) & \CSnValidDefacersHaxorID~(\CSnValidDefacersProportionHaxorID\%) & \CSnValidDefacersDefacerPro~(\CSnValidDefacersProportionDefacerPro\%) &  \CSnValidDefacersTotal~(\CSnValidDefacersProportionTotal\%)\\
Defacers with invalid reports~\textsuperscript{$\dagger$} & \CSnInvalidDefacersZoneH~(\CSnInvalidDefacersProportionZoneH\%) & \CSnInvalidDefacersOwnzYou~(\CSnInvalidDefacersProportionOwnzYou\%) & \CSnInvalidDefacersZoneXSec~(\CSnInvalidDefacersProportionZoneXSec\%) & \CSnInvalidDefacersHaxorID~(\CSnInvalidDefacersProportionHaxorID\%) & \CSnInvalidDefacersDefacerPro~(\CSnInvalidDefacersProportionDefacerPro\%) &  \CSnInvalidDefacersTotal~(\CSnInvalidDefacersProportionTotal\%)\\
\arrayrulecolor{black}
\midrule
\end{tabular}
\begin{tablenotes}
\item \tikzcmark~fully available; \dashcmark~partly available;~~$\cdotp$~not available;~~\textsuperscript{$\dagger$} duplicated defacements and defacer handles within and across different archives were unified.
\end{tablenotes}
\label{tab:active-defacement-archives}
\end{table*}

\section{Background and Related Work} \label{sec:background}
Information warfare has long been part of `hybrid' modern conflicts, especially around the control of communications~\cite{hoffman2007conflict,libiseller2023hybrid}. The enemy's ability to spread news and propaganda can be degraded by targeting crucial sites, public services, broadcast and telecom infrastructure. Censorship is often used during wartime~\cite{price1942governmental}; governments block access to global services, especially social networks and media platforms, to suppress unwanted narratives. Russia blocked news and anti-war domains when the conflict started~\cite{domainblockedbyrussia,rameshnetwork}, and lost access to foreign sites~\cite{rameshnetwork} as well as service providers~\cite{jonker2022ru}. Ukrainian users experienced degraded network performance~\cite{jain2022ukrainian}, while Ukrainian supporters tried unconventional channels such as online reviews to bypass censorship~\cite{moreno2023reviewing}. Attacks are not just online; Russian missiles hit TV towers in Kyiv in early March 2022~\cite{russiatargettvtower}.

Some associations between kinetic warfare and `nationalistic' cyberattacks have been reported. Ukrainian firms were hit by data wipers such as CaddyWiper and NotPetya~\cite{datawipingmalwareukraine2,datawipingmalwareukraine}, DDoS attacks~\cite{ukrainianunderddos,ukraineinternetproviderunderddos} and phishing campaigns~\cite{ukrainianunderphishing}; Ukraine supporters have used spam senders to distribute propaganda in Russia~\cite{ukrainespamsender} and have stolen cryptocurrency from Russian wallets~\cite{ukraineseizedcryptofromrussian}. Ukrainian universities were hacked \cite{ukrainehackwordfence}, the Ukrainian electricity grid was hit by Industroyer2~\cite{industroyer2headmind}, and the Ukrainian satellite Internet was downed~\cite{uasatelliteinternet}. Attackers self-identifying with the Anonymous movement declared a `cyberwar' on Russia~\cite{anonymouscyberwar} with attacks against Russian Ministry of Defence databases~\cite{anonymousleakdb} and state TV channels~\cite{anonymoushackstatetv}, while Killnet struck back~\cite{killnettakedown}. Russia intermittently received attacks instigated by volunteer hacktivists of the IT Army of Ukraine~\cite{itarmyukraine,foreignpolicyitarmy}.

While the security industry has reported some insights~\cite{netscout2022ua,netscout2022ru,fogofwar,microsoftreport}, empirical quantitative academic work analysing the link between armed conflicts and cybercrime has been limited. A notable report is by a Czech university's incident response team, showing negligible impact on their network after hundreds of users launched DDoS attacks against Russia for a week after the invasion~\cite{husak2022handling}.

One type of attack linked with the low-level cybercrime actors is website defacement~\cite{romagna2017hacktivism}, which accounted for around 20\% of online attacks in 2014~\cite{cyberattackstatistic2014} and is often organised into discrete campaigns~\cite{maggi2018investigating}. Attackers (or defacers) gain unauthorised access using off-the-shelf tools and simple exploits, then alter sites' appearance to demonstrate success~\cite{maggi2018investigating}. Defacers have heterogeneous developmental trajectories~\cite{van2021heterogeneity}; they are often organised in groups~\cite{perkins2022illicit} and have been using online archives~\cite{kurzmeier2020towards} as a `hall of fame' to show off their achievements to gain reputation. Defacements are mostly hobbies or pranks with greetings to peers~\cite{woo2004hackers}, but some advertise hacking tools or services to make money or display other motives such as a wish for community recognition, patriotic, religious and political views~\cite{banerjee2021using,romagna2017hacktivism}. Web defacement may cause economic harm~\cite{denning2011cyber,andress2013cyber} and has occasionally been used as a proxy for terrorist and other serious activities~\cite{holt2022examining}. Another simple type of large-scale attack linked to low-level cybercrime actors is amplified DDoS, with many attributed to DDoS-for-hire or `booter' services~\cite{krupp2017linking}. Such services abound~\cite{karami2013rent}, and off-the-shelf DDoS tools are widely available; they were tailored and provided to pro-Ukrainian volunteers early in the conflict to attack Russian infrastructure and assets.

Unlike state-sponsored activities, defacement and DDoS attacks can be systematically collected and measured with reasonable coverage. Defacements are available on online archives~\cite{kurzmeier2020towards}, while DDoS attacks can be collected through honeypots~\cite{thomas20171000,kramer2015amppot,nawrocki2023sok}. Launching these attacks with ready-made tools is straightforward for those without much technical expertise. They can be executed quickly, at scale, and have instant, noticeable effects such as altering targets' appearance, making them inaccessible, or taunting opponents with compromised sites. During wartime, the need to rapidly disseminate political messages and propaganda makes them attractive.

\section{Methods and Datasets} \label{sec:datasets}
We use several quantitative datasets collected regularly and separately, spanning 1 January to 30 June 2022;\footnote{~No further substantial changes have been observed beyond the six-month mark. We thus decided to maintain that timeframe, which is sufficient to deliver our narratives.} timestamps are normalised to UTC. To determine if the conflict has impacts resulting in different means (or mean ranks) of daily cyberattacks and hacking discussions, we separate the period into three eras; $E_1$: before the invasion, from 1 January to 24 February 2022; $E_2$: around one month immediately after the invasion, from 24 February to 31 March 2022; and $E_3$: from 1 April to 30 June 2022. We then apply unpaired statistical tests, using One-way ANOVA or Kruskal-Wallis depending on the data distribution; the null hypothesis $H_0$ is there is no significant difference between the three eras. We use post-hoc tests: Tukey-Kramer for ANOVA or Dunn's for Kruskal-Wallis to identify pairs causing changes if any. Effect sizes are measured by $\eta^2$, ranging [0, 1]; $0 \leq \eta^2 < 0.01$: no effect; $ 0.01 \leq \eta^2 < 0.06$: small effect; $0.06 \leq \eta^2 < 0.14$: medium effect; $0.14 \leq \eta^2 \leq 1$: large effect~\cite{miles2001applying}.

\para{Web Defacement Attacks}
We fully scrape the most popular active defacement archives during the period; see Table~\ref{tab:active-defacement-archives}. We started with \zoneh, the largest and most popular one (since March 2002) providing cybersecurity news and self-reported defacements along with hacking content~\cite{kurzmeier2020towards}. We then took out the most active defacers from \zoneh and investigated their online presence, including the archives on which they report attacks. Some defacers promote their attacks on Twitter and Telegram; we also looked there. We then shortlisted the five largest defacement archives by attack volume, including \ownzyou (since January 2021), \zonexsec (since May 2020), \haxorid (since November 2019), and \defacerpro (since June 2021). Smaller archives were historically active~\cite{maggi2018investigating}, but either vanished (\hackmirror and \mirrorzone) or have hosted different content (\hackcn and \mydeface). While not all compromised sites get reported, measuring trends from the most reputed archives is likely indicative. The country of defaced sites is identified based on ccTLD, IP geolocation, and geolocation of the AS hosting the sites, excluding CDNs (Appendix~\S\ref{appendix:determining-geolocation}). The defacement submission process is detailed in Appendix~\S\ref{appendix:web-defacement-submission-process}. We ensure data completeness and bypass challenges e.g., \captcha and IP blocking (Appendix~\S\ref{appendix:web-defacement-collection}). 

Further steps are performed to enhance data reliability. First, many on-hold submissions are valid but were never verified; we perform a semi-automatic validation using the messages left on defaced pages (see Appendix~\S\ref{appendix:validating-defacement}). Second, submissions may be reported to multiple archives to broaden their visibility. We de-duplicate across and within archives by hashing their content (see Appendix~\S\ref{appendix:unifying-defacement-defacer}). Third, as `notifier' can be arbitrary, typos can give a single attacker multiple identities; we correct typos across all archives using handles' similarity and messages left on defaced pages (see Appendix~\S\ref{appendix:unifying-defacement-defacer}). In total, \CSnDefacementUnifiedValidStaffVerificationTotal~reports were verified by the archives, \CSnDefacementUnifiedValidAutoVerificationTotal~were automatically validated by us and a further \CSnDefacementUnifiedValidManualVerificationTotal~were validated semi-automatically; \CSnDefacementRawMinusUnifiedTotal~(\CSnDefacementRawMinusUnifiedProportionTotal\%) duplicate reports are merged across all archives. Of the remaining \CSnDefacementUnifiedTotal~reports, we analyse \CSnDefacementUnifiedValidTotal~validated submissions (\CSnDefacementUnifiedValidProportionTotal\%, around 1\,500 per day). Of these, \CSnDefacersRawTotal~defacer handles are also unified to \CSnTotalHandlesTotal.

\para{UDP Amplification DDoS Attacks} We use \CSnDDoSAttacksRounded~DDoS attack records gathered by a honeypot network emulating protocols vulnerable to reflected UDP attacks~\cite{thomas20171000}. A flow of packets is considered to be an attack if any sensor observes at least five packets for the same victim IP or IP prefix, and the attack is deemed to last from the first packet until the last packet preceding 15 minutes without further packets. In 2022, the median number of honeypots contributing data was \hsSensorsDayTwentyTwentyTwoMedian, \hsSensorsDayTwentyTwentyTwoCINF; the median number of observed attacks per week was \hsObsAttacksWeekTwentyTwentyTwoMedian, \hsObsAttacksWeekTwentyTwentyTwoCINF\ and on IP prefixes of \hsPrefixAttacksPerWeekTwentyTwentyTwoMedian, \hsPrefixAttacksPerWeekTwentyTwentyTwoCINF; the median attack duration was \hsAttackDurationTwentyTwentyTwoFiftyP\ minutes, while the maximum was \hsAttackDurationTwentyTwentyTwoOneHundredP\ minutes. The country of victims is identified by geolocation of the IP address and AS hosting it, excluding CDNs (see Appendix~\S\ref{appendix:determining-geolocation}).

\para{Underground Forum Discussions} Online forums are structured around subforums containing threads with multiple posts. To assess changes in discussion topics within the hacking community, we use a snapshot of the most popular hacking forum, \hackforums from the \crimebb~dataset~\cite{pastrana2018crimebb}. The forum is a place for users to learn about attacks and trade cybercrime tools and services. Many are low-level actors; however, some have been prosecuted for cybercrime-related activities~\cite{pastrana2018characterizing}. We extract all \CSnHFRUUAFinalThreads~threads within the six months consisting of at least one post with the keywords `Russia' and/or `Ukraine' (case-insensitive): \CSnHFRUThreads~related to Russia, \CSnHFUAThreads~related to Ukraine, in which \CSnHFRUUAOverlapsThreads~related to both. We then use all \CSnRUUAHighlyRelevantPosts~posts from \CSnRUUAHighlyRelevantThreads~highly relevant threads -- those with titles directly having the keywords. For the rest \CSnRUUALowerRelevantThreads~less-relevant threads, we count \CSnRUUALowerRelevantPosts~posts directly consisting of the keywords. In total, \CSnRUUAAllRelevantPosts~relevant posts made by \CSnRUUAAllRelevantUsers~active forum users are analysed.

\begin{figure*}[t]
    \centering
    \includegraphics[width=\textwidth]{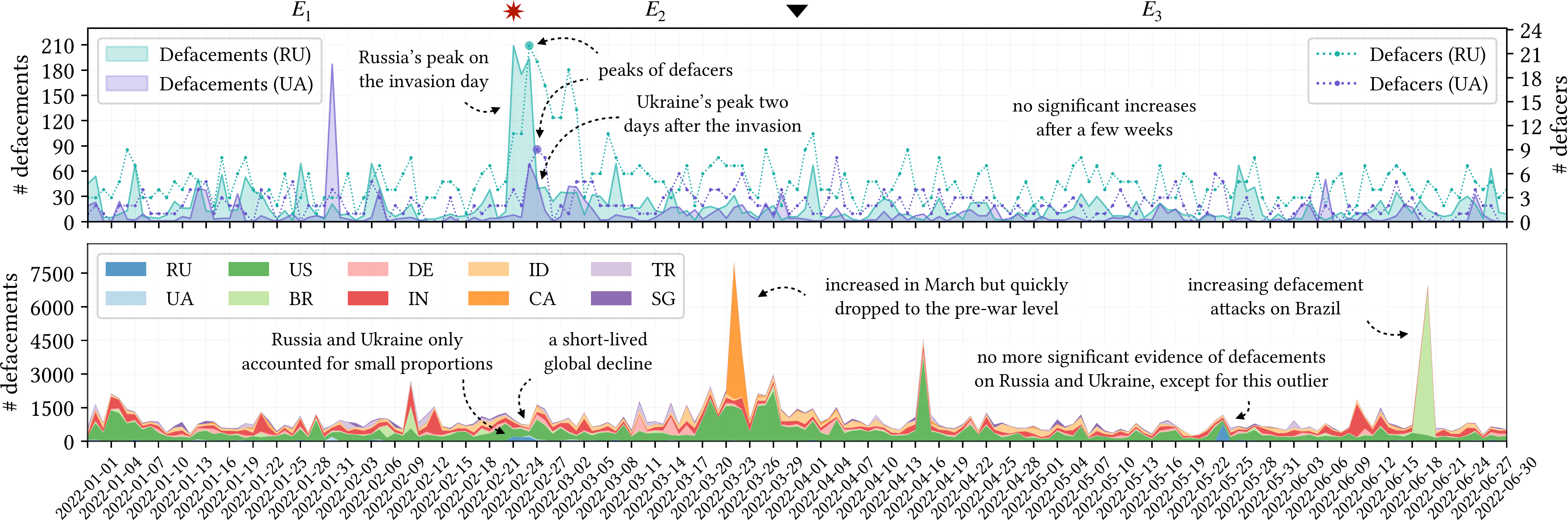}
    \caption{Number of defacements and defacers per day in the Russia-Ukraine scale (top) and the global scale (stacked, bottom).}
    \label{fig:defacements-volume}
\end{figure*}
\begin{figure}[t]
    \centering
    \includegraphics[width=0.45\textwidth]{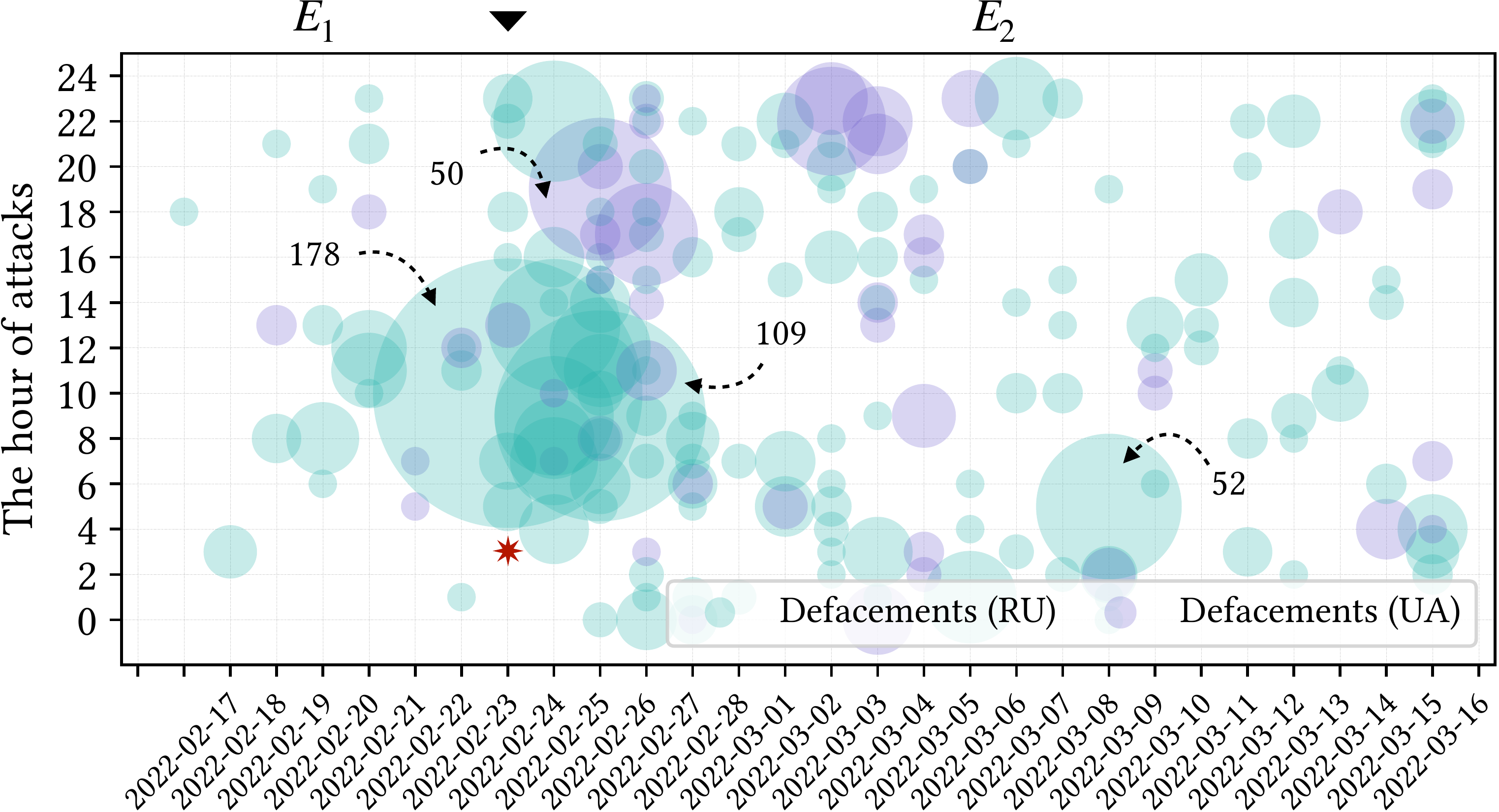}
    \caption{Number of defacements hitting Russia and Ukraine by hour around the invasion day (marked with the red star).}
    \label{fig:defacements-ru-ua-hourly}
\end{figure}

\para{Volunteer Hacking Discussions} Two days after the invasion, the Ukrainian government called on pro-Ukraine `hacktivists' to join the IT Army of Ukraine, which was stood up in an ad-hoc manner~\cite{soesantoitarmy} to support the war effort~\cite{uagovernmentcreatingitarmy,theguardianitarmy}. The most tangible outcome is a public Telegram channel used mainly to recruit and encourage volunteers to spread political propaganda and co-ordinate disruptive efforts against Russia. We confirmed with a Ukrainian government source that it is the official channel used for communication amongst Ukrainian civilians, with messages being forwarded to other unofficial satellite groups with far fewer subscribers. 

The group, with over 200k subscribers, promotes lists of Russian targets (both in Ukrainian and English) most mornings with URLs and IP addresses posted on their channels. They encourage using various attack vectors to disrupt communication and financial systems by hitting banks, businesses, government, and logistics~\cite{itarmyukraine}. They provide guides and tools to launch attacks e.g., quickly fetching daily targets and granting access to individuals' cloud resources for later coordinated attacks, claiming ordinary Russians have seen effects when they hit banks, exchanges~\cite{forbesrussianbankddos}, and cinemas~\cite{itarmnycinemas}.

We believe the involved `volunteer hacktivists' are mostly low-level actors, as much of their activity depends on tools provided by the group. We collect \CSnTelegramMessages~announcements with \CSnTelegramReplies~replies and \CSnTelegramEmojisRounded~emoji reactions posted in the channel from its inception until 30 June 2022 using Telethon, which interacts with official Telegram APIs to fully capture messages and metadata. We then used regular expressions to extract promoted IP addresses and domains; subdomains such as www.xyz.ru and smtp.xyz.ru are combined. Besides Russian and Belarusian domains (.ru, .su, .by), top-level domains (e.g., .tv, .com) are also targeted. URL shorteners (e.g., goo.gl) and online services (e.g., youtube.com) are excluded, resulting in \CSnTelegramPromotedTargets~targets: \CSnTelegramPromotedTargetsIPs~IP addresses (\CSnTelegramPromotedTargetsIPsProportion\%) and \CSnTelegramPromotedTargetsDomains~domains (\CSnTelegramPromotedTargetsDomainsProportion\%).

\section{The Evidence from Web Defacements} \label{sec:defacements-evidence}
We measure the dynamic of defacement, both on the global and Russia-Ukraine scales. Figure~\ref{fig:defacements-volume} shows the number of defacements per day as the conflict progressed. Figure~\ref{fig:defacements-ru-ua-hourly} breaks down changes by hour for the most active four-week period from 17 February 2022.

\para{The Russia-Ukraine Scale} Website defacements targeting Russia immediately peaked on the invasion day at \CSdefacementRUAfterWar~(\CSdefacementProportionRUAfterWar\% of all defacements on that day, while it was \CSdefacementProportionRUBeforeWar\% the day before). The first big wave was at around 10 AM (7 hours after the invasion) with 178 attacks caused by a single defacer, followed by smaller waves on the same day. Two follow-up waves occurred at 1 PM on 25 February and 9 AM on 26 February with 43 and 109 attacks, respectively. The number of defacers targeting Russia peaked 2 days later: while only 11 defacers accounted for the peak on 24 February, there were 22 on 26 February. No notable change in defacements hitting Ukraine was seen on the invasion day, but a peak of \CSdefacementUATwoDayAfterWar~attacks occurred 2 days later (\CSdefacementProportionUATwoDayAfterWar\% of all defacements on that day, while it was \CSdefacementProportionUAOneDayAfterWar\% the day before). The largest wave was at around 7 PM on 26 February (50 attacks), followed by medium waves at 5 PM on 27 February (26 attacks) and 10 PM on 3 March (29 attacks). Defacers targeting Ukraine peaked with 9 on 27 February, 1 day after the largest wave.

There was a spike of 771 web defacements by five defacers targeting Russia on 25 May. Of these, 764 were claimed by a single defacer compromising a server hosting 760 sites. This outlier appears to be unique; it was removed from the graph for better visualisation. The peak of 187 web defacements hitting Ukraine on 1 February 2022 by four defacers did not have a single cause and did not lead to a sharp increase in the number of defacers in the following days.

For both Russia and Ukraine, Kruskal-Wallis tests suggest a statistically significant difference in the number of web defacements and defacers per day through $E_1$, $E_2$, and $E_3$, see Table~\ref{tab:statistical-significance-web-defacements} (Appendix~\S\ref{appendix:significance-impact-levels}). Post-hoc analysis indicates a significant difference between the pre-invasion ($E_1$) and one-month-post-invasion ($E_2$) periods. $p\langle E_2, E_3 \rangle$ is also significant, but \textit{not} $p\langle E_1, E_3 \rangle$, suggesting the situation returned to the pre-war levels after the second era. The effect sizes $\eta^2$ are all between medium and large, ranging [0.06, 0.14].

\para{The Global Scale} The number of web defacement attacks targeting Russia and Ukraine appears to be trivial when set against the global scale. Among \CSnDefacementUnifiedValidTotal~analysed web defacements, only \CSnDefacementUnifiedValidRUUA~(\CSnDefacementUnifiedValidRUUAProportion\%) targeted the two countries (\CSnDefacementUnifiedValidRU~for Russia and \CSnDefacementUnifiedValidUA~for Ukraine). The top 10 countries account for \CStotalDefacementProportionTopCountries\% of all web defacements; sites hosted in the US by global-scale vendors (e.g., Cloudflare, Amazon) are excluded (see Appendix~\S\ref{appendix:determining-geolocation}), but the US still consistently suffers the majority of web defacements. Since January 2022, the US accounts for \CStotalDefacementProportionUS\% of defacement attacks, followed by India with \CStotalDefacementProportionIN\% and Indonesia with \CStotalDefacementProportionID\%, while Russia and Ukraine only account for small proportions: \CStotalDefacementProportionRU\% and \CStotalDefacementProportionUA\%, respectively.

There was a short-lived decline in defacement attacks worldwide on the invasion day (from around 1\,400 to 1\,000), while it peaked for Russia from nearly zero to \CSdefacementRUAfterWar~(\CSdefacementProportionRUAfterWar\% of all defacements). This suggests a genuine change in the way defacers chose their targets, precipitated by the war. The US is consistently the largest target but only accounts for \CSdefacementProportionUSAfterWar\% on that day. During the last two weeks of March 2022, the number of defacements significantly increased at the global scale, with many defacements targeting the US. However, much like the Russia-Ukraine scale, the effect lasted for only a few weeks. The unusual peaks against Brazil happened in late June (also for DDoS attacks, see Section~\S\ref{sec:ddos-attacks-evidence}), without a clear explanation.

The patterns seen from the Kruskal-Wallis and post-hoc tests in the Russia-Ukraine scale do not apply for most top countries, see Table~\ref{tab:statistical-significance-web-defacements} (Appendix~\S\ref{appendix:significance-impact-levels}). ANOVA/Kruskal-Wallis tests on the defacement and defacer counts are not \textit{both} significant for Brazil, Germany, India, and Singapore; no significant changes are seen between the pre-invasion and one-month-post-invasion eras $\langle E_1, E_2 \rangle$ for Canada and Turkey. Indonesia has a similar pattern of post-hoc tests, but one of the effect sizes is small. The only country following a close pattern is the US with medium effect sizes, yet Section~\S\ref{sec:ddos-attacks-evidence} will show this did not hold for DDoS attacks on the US.

The evidence above suggests a genuine increase of website defacements against the two countries shortly after the invasion, standing out significantly from other top countries. Russia was the first to be hit at scale, followed by Ukraine a few days later. However, this effect was fairly short-lived for both countries, lasting for only a few weeks before returning to pre-war levels, presumably as defacers ran out of targets or just lost interest in carrying out attacks. The number of defacers involved was small, but for a while, they turned from indiscriminate to more targeted attacks.

\para{Defacement Motives} The conflict caught the attention of existing defacers, who performed many attacks against other countries but not Russia and Ukraine until just after the invasion, suggesting their choice of targets was influenced. We also found some `new faces' e.g., the second most active defacer targeting Russia after the war began first appeared in mid-February, peaked on the invasion day, stayed significant for three days then declined quickly. While some minor players at the global scale made a significant contribution to the rise in attacks on Russia and Ukraine, the three most active defacers globally made a trivial contribution (less than 10) against either country. We do not verify findings on their general motives (see~\S\ref{sec:background}), but to gain conflict-related insights we analyse the contents of \CSnDefacementUnifiedValidRU~defacements targeting Russia and \CSnDefacementUnifiedValidUA~hitting Ukraine. 

We annotate motives based on \CSnDefacementUnifiedValidRUUAGroupedByMessage~unique messages left on the defaced pages. We consider a political sentiment and mark it as supporting Russia/Ukraine if a support/objection is expressed e.g., `\textit{We stand with Ukraine!}'. We mark messages consisting of defacer signatures e.g., `\textit{Hacked by ABC}' without clear motives, or just greetings to peers as being for self-aggrandisement. Messages advertising hacking tools and services or asking for ransom are marked financially motivated e.g., `\textit{Contact me for shells}'. We label messages expressing favourite mottos or moods as self-expression e.g., `\textit{Not much I want, hope my life will be better}', and exclude \CSnDefacementWarRelatedNoneDeterministicMotives~messages (\CSnDefacementWarRelatedNoneDeterministicMotivesProps\%) containing empty or random messages.

We find diverse motives, but despite targeting Russia and Ukraine, most messages do not refer to the conflict. \CSnDefacementWarRelatedFunMotives~(\CSnDefacementWarRelatedFunMotivesProps\%) were for self-aggrandisement, \CSnDefacementWarRelatedSelfExpressionMotives~(\CSnDefacementWarRelatedSelfExpressionMotivesProps\%) self-expression,  \CSnDefacementWarRelatedNationalisticConflictsMotives~(\CSnDefacementWarRelatedNationalisticConflictsMotivesProps\%) related to other conflicts (such as Israel-Palestine), \CSnDefacementWarRelatedPatriotiMotives~(\CSnDefacementWarRelatedPatriotiMotivesProps\%) related to patriotism, and \CSnDefacementWarRelatedFinanceMotives~(\CSnDefacementWarRelatedFinanceMotivesProps\%) were financially motivated (mainly from the two most active defacers globally who did not change their targeting due to the conflict). Some defacers did leave conflict-related messages: \CSnDefacementWarRelatedProUAMotives~(\CSnDefacementWarRelatedProUAMotivesProps\%) supporting Ukraine, roughly 2.8 times higher than those supporting Russia at \CSnDefacementWarRelatedProRUMotives~(\CSnDefacementWarRelatedProRUMotivesProps\%). Notably, some defacers supported Russia, yet also defaced Russian sites, saying they wished to alert and help secure the systems (\CSnDefacementWarRelatedProRUMotivesButTargetRU~attacks) -- `\textit{I have secured this domain, I love Russia}', was a message the third most active pro-Russia defacer left on a Russian website. Likewise, other defacers supported Ukraine yet defaced Ukrainian sites (\CSnDefacementWarRelatedProUAMotivesButTargetUA~attacks) e.g., `\textit{Hello Volodymyr Zelensky, I'm sorry to hack your site. I just wanted to tell you that people need a president like you. We support Ukraine}'. Such signatures are likely intentionally war-related as Russia and Ukraine were not frequently targeted before.

\begin{figure*}[t]
    \centering
    \includegraphics[width=\textwidth]{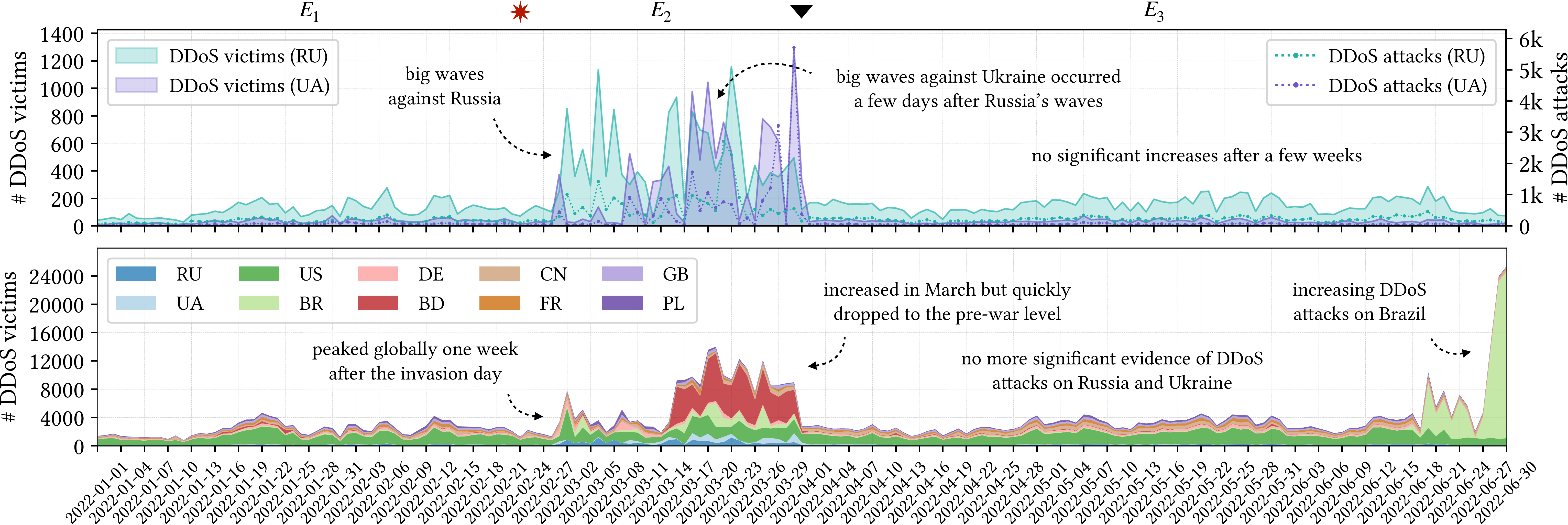}
    \caption{Number of DDoS attacks and victims per day in the Russia-Ukraine scale (top) and global scale (stacked, bottom).}
    \label{fig:ddos-attacks-volume}
\end{figure*}
\begin{figure}[t]
    \centering
    \includegraphics[width=0.45\textwidth]{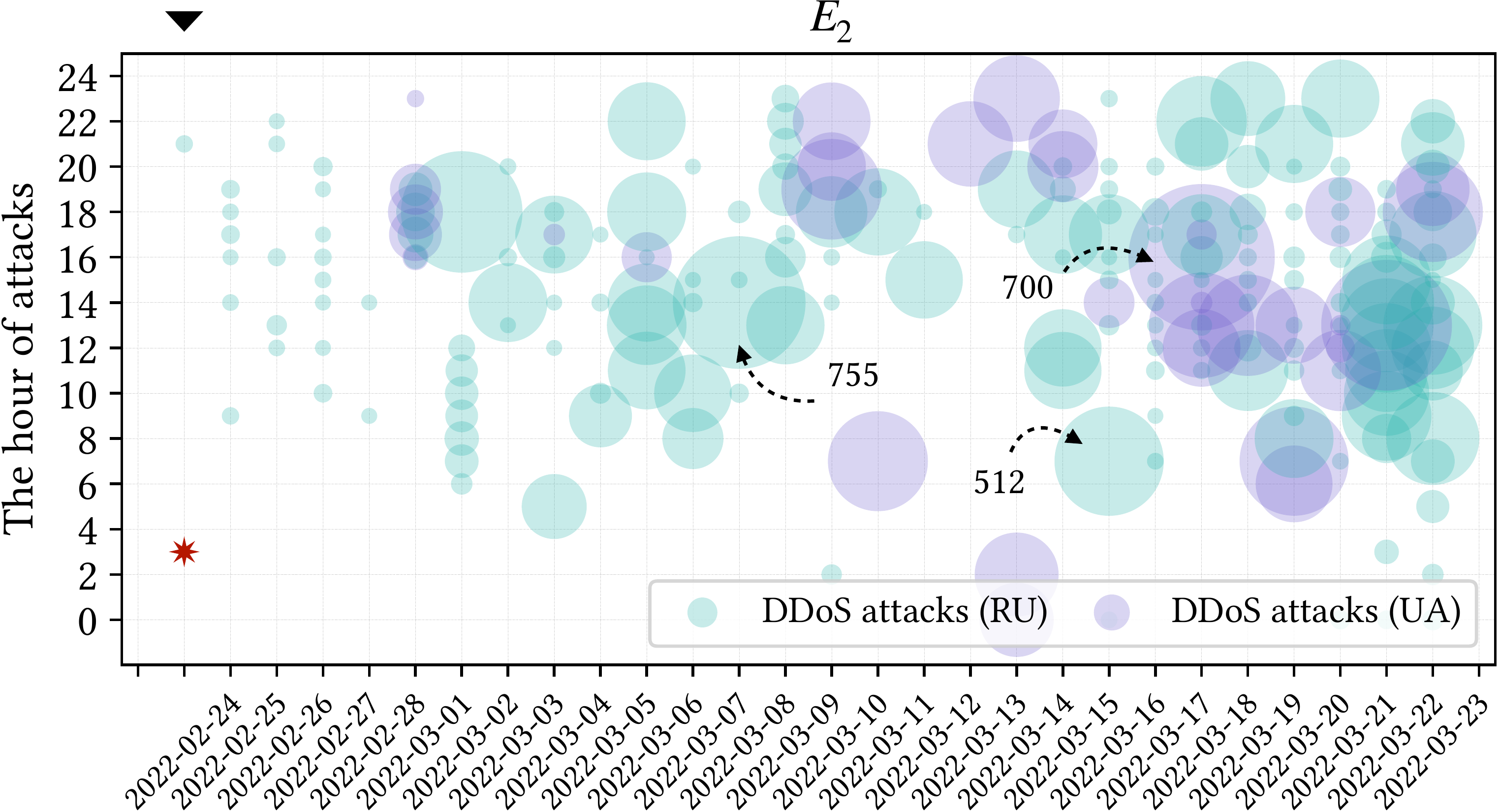}
    \caption{Number of DDoS attacks on Russia and Ukraine by hour around the invasion day (marked with the red star).}
    \label{fig:ddos-attacks-ru-ua-hourly}
\end{figure}

\section{The Evidence from DDoS Attacks} \label{sec:ddos-attacks-evidence}
We now examine if there were also significant changes in DDoS attack volumes targeting Russia and Ukraine after the conflict. Figure~\ref{fig:ddos-attacks-volume} shows the number of DDoS attacks in both Russia-Ukraine and global scales over the three eras, while Figure~\ref{fig:ddos-attacks-ru-ua-hourly} shows their changes by hour during the most active four weeks from 24 February.

\para{The Russia-Ukraine Scale} DDoS attacks lagged defacement by about a week, but occurred in higher volumes and lasted longer; most happened after 7 AM. The number of both DDoS attacks and victims targeting Russia first increased on 2 March (six days after the invasion) with 851 victims, 511 of them at around 6 PM. The attacks peaked four days later with 1\,137 victims. High activity levels continued through 23 March, with the biggest wave occurring at around 2 PM on 8 March with 755 victims. Smaller waves continued regularly during the next few weeks. Regarding DDoS attacks hitting Ukraine, significant waves started around a week after Russia's first big wave (some small spikes targeting Ukraine before Russia were insignificant) with the first notable spike on 10 March having 526 victims, then became prevalent during two weeks from 18 to 31 March: big waves were on 18 March at around 12 PM, 1 PM and 4 PM with 257, 476, and 700 victims, respectively. Other big and medium waves lasted until the end of March, with the biggest peak on 31 March when 1\,296 victims were hit. The increased volume only continued for about a month before declining sharply.

Kruskal-Wallis tests suggest statistically significant changes between the daily number of DDoS attacks and victims through the three eras for both Russia and Ukraine, much like with defacements; see Table~\ref{tab:statistical-significance-ddos-attacks} (Appendix~\S\ref{appendix:significance-impact-levels}). Post-hoc analysis shows high significance levels of $\langle E_1, E_2 \rangle$ and $\langle E_2, E_3 \rangle$, suggesting notable changes between the pre-invasion vs. one-month-post-invasion periods, and the one-month-post-invasion periods vs. the period after that. The main difference between Russia and Ukraine is that the situation for Ukraine returned to pre-invasion levels after one month i.e. $\langle E_1, E_3 \rangle$ is not significantly different, while we still see some difference with Russia. Indeed, the number of DDoS attacks hitting Russia was still slightly higher than before the invasion (see Figure~\ref{fig:ddos-attacks-volume}). The effect size is large for Russia, while it is medium for Ukraine.

\begin{figure*}[t]
    \centering
    \includegraphics[width=\textwidth]{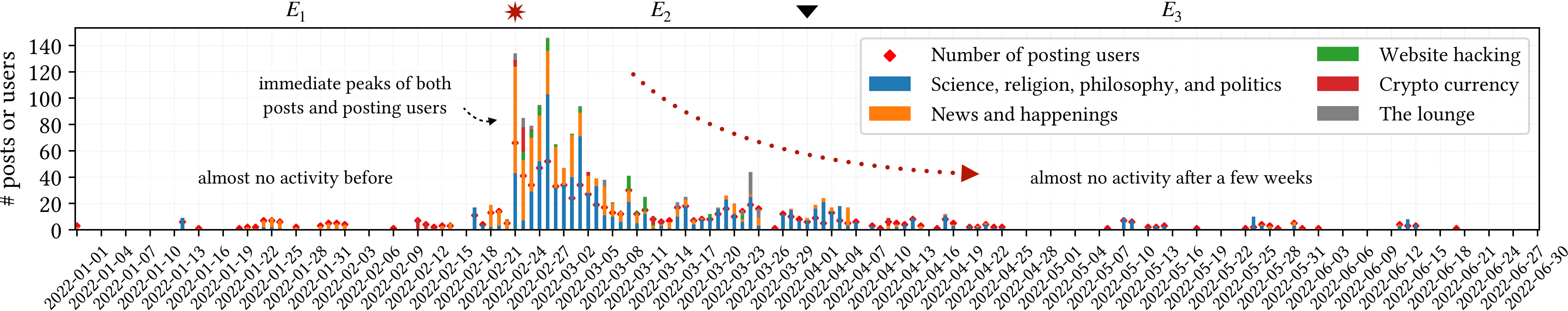}
    \caption{Number of daily posts and posting users on Hack Forums mentioning Russia and/or Ukraine (the top 5 subforums).}
    \label{fig:hacking-discussion-about-two-countries}
\end{figure*}

\para{The Global Scale} We again see concentrations in DDoS attacks, with the top 10 countries accounting for \CStotalDDoSVictimProportionTopCountries\% of all victims. The US still dominates (\CStotalDDoSVictimProportionUS\%), followed by Brazil (\CStotalDDoSVictimProportionBR\%) and Bangladesh (\CStotalDDoSVictimProportionBD\%). Ukraine took \CStotalDDoSVictimProportionUA\%, while Russia lies 8\textsuperscript{th} at \CStotalDDoSVictimProportionRU\%. Our DDoS and defacement datasets show some correlations. Three of the top 10 countries for defacements are also in the top 10 for DDoS targets: the US, Germany, and Brazil. As with defacements, DDoS attacks rose globally during the last two weeks of March 2022 with large numbers targeting Bangladesh, but this effect is insignificant; see Table~\ref{tab:statistical-significance-ddos-attacks} (Appendix~\S\ref{appendix:significance-impact-levels}). The unusual peaks of both defacements and DDoS against Brazil in late June are notable; Brazil is often ranked among the top cybercrime hubs worldwide~\cite{lusthaus2020mapping}, yet we lack a convincing causality. A similar peak observed on the Russia-Ukraine scale can also be seen at the global scale following the invasion. As with defacements, DDoS attacks thrived on a global scale in March, yet quickly returned to the previous levels after a few weeks with no lasting global changes and no significant evidence of further waves, whether targeting Russia or Ukraine.

The patterns seen from the Kruskal-Wallis and post-hoc tests in the Russia-Ukraine scale do not apply for most top countries, see Table~\ref{tab:statistical-significance-ddos-attacks} (Appendix~\S\ref{appendix:significance-impact-levels}). Kruskal-Wallis tests are not \textit{all} significant for the US, Bangladesh, and Poland, despite the US accounting for the largest number of attacks and there was a visual increase for Bangladesh (as the tests compare mean ranks instead of means). No significant changes are seen between the pre-invasion and one-month-post-invasion periods $\langle E_1, E_2 \rangle$ for France and the UK. Brazil and Germany have a similar phenomenon of post-hoc tests in the Russia-Ukraine scale, yet one effect size is small. China is the only country following that phenomenon with large effect sizes; the main difference is that the changes in $p\langle E_2, E_3 \rangle$ are not significant.

As with defacements, the evidence above suggests a genuine increase in DDoS attacks targeting Russia and Ukraine as the conflict began, standing out significantly from most top countries. Russia was still the first to be hit at scale, followed by Ukraine shortly after. The outbreak of both defacement and DDoS attacks on Russia and Ukraine was significant and timely, but fairly short-lived. While defacements returned to previous levels after a few weeks, DDoS attacks on Russia remained marginally higher than pre-war levels.

\begin{figure}[t]
    \centering
    \includegraphics[width=0.45\textwidth]{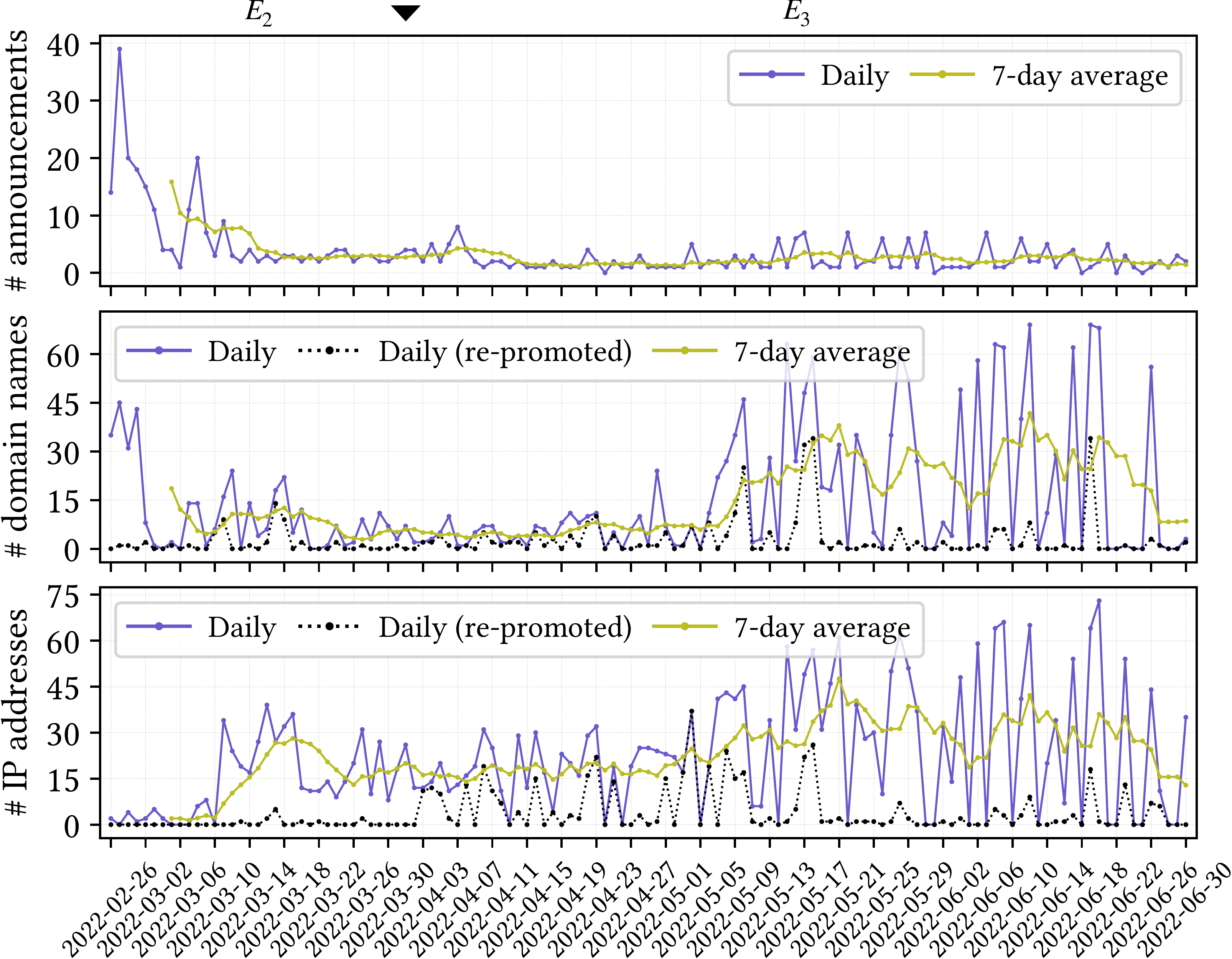}
    \caption{The number of announcements and (re-promoted) targets in the IT Army of Ukraine Telegram channel by day.}
    \label{fig:telegram-promoted-targets}
\end{figure}

\section{Hacking Community Reactions} \label{sec:community-discussions}
\para{\hackforums Discussions} There was an immediate surge of posts on \hackforums mentioning the two countries after the invasion, from near zero to over 120 per day; see Figure~\ref{fig:hacking-discussion-about-two-countries}. \CSnRUUARelatedPostsSignificanceTestMethod~tests confirm the significance $H(\CSnRUUARelatedPostsSignificanceTestDoF) = \CSnRUUARelatedPostsSignificanceTestStat, \CSnRUUARelatedPostsSignificanceTestPValue$, with a large effect size~$\eta^2=\CSnRUUARelatedPostsEffectSize$; pairwise post-hoc tests for  $\langle E_1, E_2 \rangle$ and $\langle E_2, E_3 \rangle$ are both significant ($\CSnRUUARelatedPostsPostHocPValueTWOTHREE$), but not $\langle E_1, E_3 \rangle$ ($\CSnRUUARelatedPostsPostHocPValueONETHREE$). The number of posting users shows a similar story: \CSnRUUARelatedUsersSignificanceTestMethod~test reports $H(\CSnRUUARelatedUsersSignificanceTestDoF) = \CSnRUUARelatedUsersSignificanceTestStat, \CSnRUUARelatedUsersSignificanceTestPValue$ with a large effect size~$\eta^2=\CSnRUUARelatedUsersEffectSize$; pairwise post-hoc tests for  $\langle E_1, E_2 \rangle$ and $\langle E_2, E_3 \rangle$ are both significant ($\CSnRUUARelatedUsersPostHocPValueTWOTHREE$), but not $\langle E_1, E_3 \rangle$ ($\CSnRUUARelatedUsersPostHocPValueONETHREE$). This fits the evidence seen with defacement and DDoS attacks: both posting activity and users returned to the pre-war level after a few weeks, presumably as users quickly lost interest and moved on to other discussion topics. 

This posting volume is tiny when set against the 62M-post size of \hackforums, showing trivial contributions of the Russia-Ukraine discussions to the overall landscape (as with the previous evidence seen from defacement and DDoS attacks). These posts are centralised: \CSnHFRUUANPostsTopCategoriesProportion\% belongs to the top 5 popular subforums. Ranked 1\textsuperscript{st} is \textit{`science, religion, philosophy, and politics'}, accounting for \CSnHFRUUANPostsTopCategoriesNumberOneProportion\%; ranked 2\textsuperscript{nd} is \textit{`news and happenings'} with \CSnHFRUUANPostsTopCategoriesNumberTwoProportion\%; \textit{`website hacking'} ranked 3\textsuperscript{rd}, followed by \textit{`crypto currency'}, then general chats. We see some \textit{`news and happenings'} posts in the past, but mostly no \textit{`science, religion, philosophy, and politics'} posts until the invasion.

\begin{figure}[t]
    \centering
    \includegraphics[width=0.45\textwidth]{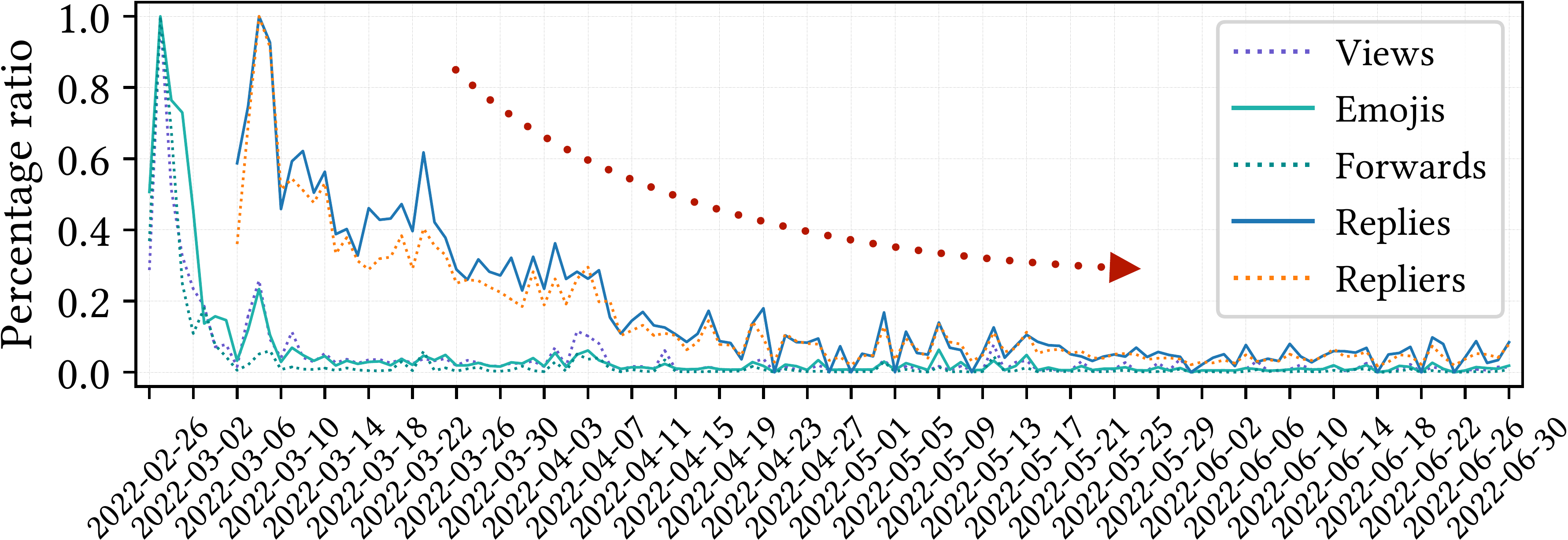}
    \caption{Levels of user engagement daily in the IT Army of Ukraine Telegram channel. Values are min--max normalised.}
    \label{fig:telegram-community-reactions}
\end{figure}

\para{Targets Promoted by the IT Army of Ukraine} Many announcements and targeted domains were posted in the first two weeks after the invasion, beginning on 26 February, peaking on 27 February with 40 announcements and 45 domains promoted (IP addresses were not regularly included until later), see Figure~\ref{fig:telegram-promoted-targets}. Yet they quickly declined to consistently less than 10 per day after two weeks with some days (e.g., 24 and 26 April) having no posts. The number of subscribers also dropped from 300k to around 160k in October 2023.

While the number of announcements dropped, the number of targets steadily increased, particularly in May and June 2022 with multiple-target posting. Activities were unstable at that time; targets got promoted less frequently and occasional days had no targets. Targets were mostly fresh in the first two weeks, but then a considerable proportion got re-promoted on multiple days e.g., all advertised IP addresses and most domains were re-posted during 4--6 May. Along with frequent zero-target days, this suggests the group might run out of new targets or get bored with finding them.

Community reactions and engagement tell much the same story as with DDoS and defacement attacks (see Figure~\ref{fig:telegram-community-reactions}). While more targets were promoted in May and June, volunteers appeared to have largely lost interest, despite their intense activity in the first few weeks. The decline in reaction was consistent across all engagement types: views, emojis, forwards, and replies. Older announcements may have more time to accrue views as people scroll up the channel, but the emojis, forwards, and replies require user intent. We believe the figures reflect a genuine decline in user engagement over time. 

The group first provided instructions about tools and guidance to carry on attacks against Russian payment systems on 9 March (two weeks after the invasion), attracting high levels of engagement: 240k views, 2.6k emojis, 1.2k forwards, and 421 replies from 197 users. The next was on 1 April: while the number of replies and forwards was similar to the first, other kinds roughly halved. From mid-May to late June, instructions were posted four more times, yet users were around four times less engaged than the first in March, indicating a loss of interest despite the operator's intensive efforts.

\begin{figure}[t]
    \centering
    \includegraphics[width=0.45\textwidth]{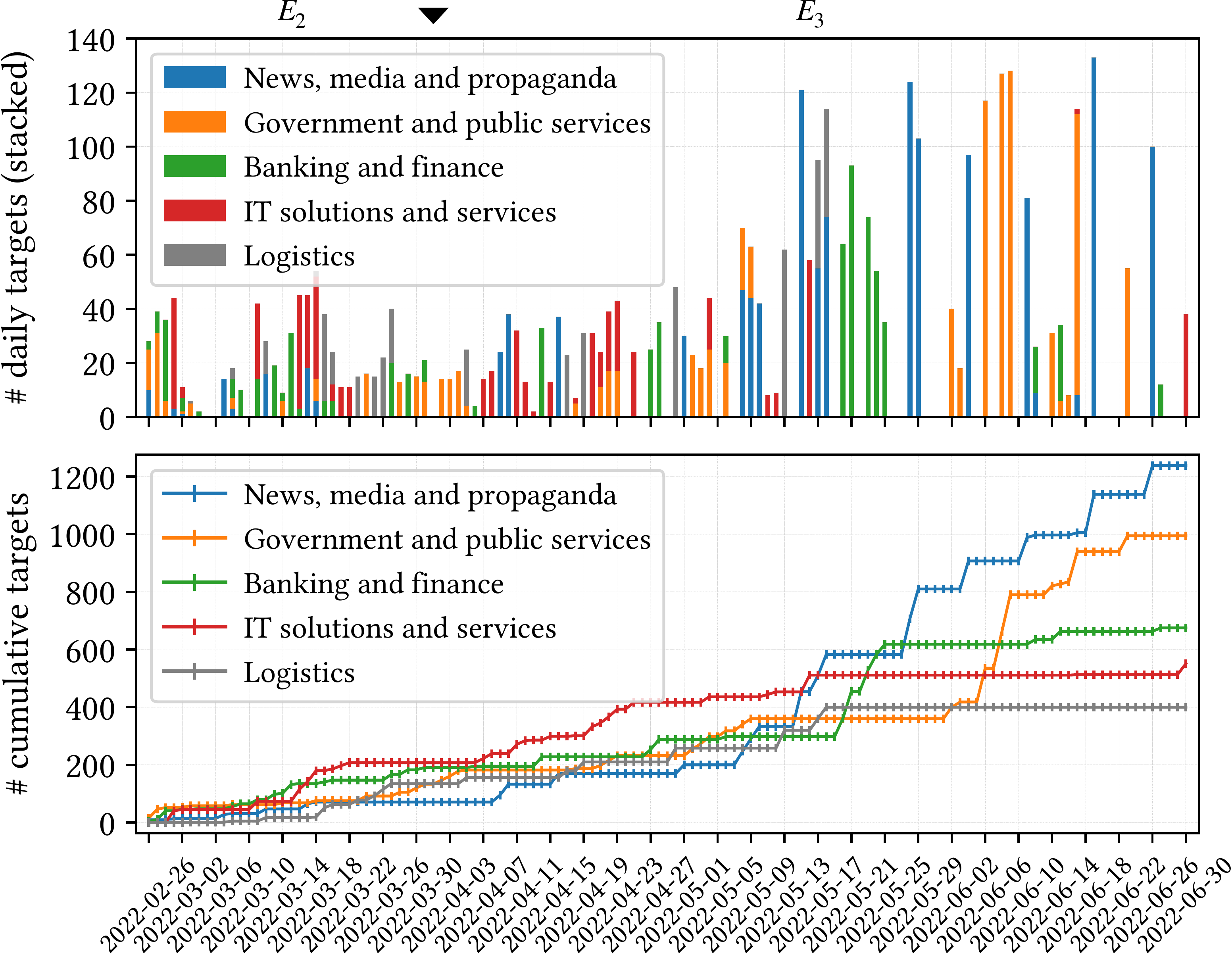}
    \caption{Number of daily stacked targets (top) and cumulative targets (bottom) being promoted in the top 5 categories.}
    \label{fig:target-categories}
\end{figure}

\para{Target Selection} Targets were often themed, patterning around particular weekdays e.g., online news and propaganda, food delivery, and entertainment are often attacked at weekends to maximise impact as people spend more time online. Themes were also occasionally set with re-promoted old targets, leading to wide variations in the number of new targets, particularly from May onwards (zero on some days, see Figure~\ref{fig:telegram-promoted-targets}). Subscribers can suggest new targets, but group organisers post most of them. During the last two months, they often re-promoted targets linked to old posts, which could be as simple as `\textit{we continue to work with yesterday's targets}', suggesting that they might have some difficulties in hunting new targets.

We use categories linked with targets by default; when unavailable, we rely on root domains e.g., .tv and .gov are likely news and government sites. Categories of generic domains (e.g., .net, .com) are identified by direct visits (via Russian IP relays) or querying Internet archives if they are down. Some targets were indeed down while previously active, suggesting attacks might have succeeded e.g., ksrf.ru (the Constitutional Court of the Russian Federation) was down for a while, and data.gov.ru was both defaced and DDoSed.

Categories vary, yet five dominate \CStopCategoriesProportion\% of all targets, see Figure~\ref{fig:target-categories}. \textit{`News, media and propaganda'}, including TV broadcasting, has been consistently promoted since the war began but only became the most common one in May when it overtook \textit{`IT solutions and services'}. \textit{`Government and public services'}, which includes military, state-owned websites, and public services for civilians such as parking and lighting (including governments imposed on occupied territories) has also been regularly targeted throughout, but they only grew rapidly towards the end of the period, making this the second most common category overall. \textit{`Banking and finance'} ranked 3\textsuperscript{rd}, including banks, stock exchanges, electronic payment, accounting, credit services, credit card providers, trading, bidding, investment platforms, funding agencies, and the microfinance industry. \textit{`IT solutions and services'} ranked 4\textsuperscript{th}, including software solutions supporting governments, digital signature and information security services such as DDoS-Guard. This sector was actively promoted early on but was targeted far less thereafter. \textit{`Logistics'} ranks 5\textsuperscript{th}, including airlines and aviation, travel, shipping, and food delivery. Other categories by popularity include (online) marketplaces and stores (e.g., job markets, real estate, e-commerce, drug stores, app stores), manufacturers and trading (e.g., military footwear, shoes, wood and roofing materials), education, insurance, telecommunications (e.g., Internet providers), businesses and state companies (e.g., energy and steel manufacturers), online forums, entertainment (e.g., cinemas), and non-governmental organisations.

\para{Crossover with Observed Attacks} The IT Army of Ukraine maintains a dashboard of targets' status, claiming many are down due to their actions. To find whether the attacks involved reflected DDoS or defacement, we correlate our attack records with promoted targets since the Telegram group started. We consider a defacement overlap when either its URL or IP address matches promoted targets, while for DDoS attacks, only IP addresses are used. There was very little overlap with website defacement attacks: among \CSnTelegramPromotedTargets~targets promoted by the group, there are only \CSnTelegramPromotedTargetsOverlapsWithDefacements~valid matches (\CSnTelegramPromotedTargetsOverlapsWithDefacementsProportion\%), including \CSnTelegramPromotedTargetsOverlapsWithDefacementsDomainOnly~domain matches (\CSnTelegramPromotedTargetsOverlapsWithDefacementsDomainOnlyProportion\%) and \CSnTelegramPromotedTargetsOverlapsWithDefacementsIPOnly~IP matches (\CSnTelegramPromotedTargetsOverlapsWithDefacementsIPOnlyProportion\%). Notably, no overlaps occur on the day targets are promoted, suggesting that defacers chose their targets themselves independently (by scanning sites based on ccTLDs); these targets are largely unimportant and irrelevant to the conflict. For DDoS attacks, we observe \CSnTelegramPromotedTargetsOverlapsWithDDoS~(\CSnTelegramPromotedTargetsOverlapsWithDDoSProportion\%) total overlaps among \CSnTelegramPromotedTargetsIPs~promoted IP addresses, which is considerable. Unlike defacements, some are executed the same day they are promoted; we find many same-day overlaps in late March, early April, and during May, peaking on 19 March 2022 with 22 victims overlapping. However, the crossover dropped quickly, becoming less frequent from late May while many new targets were still actively advertised. This suggests a loss of interest by volunteers in attacking targets promoted by the group.

\section{Concluding Remarks} \label{sec:discussion}
The role of the low-level cybercrime actors studied in this paper amounts to essentially trivial acts of solidarity and opportunistic competition. Their primary impact is probably to disseminate political propaganda, with little measurable evidence to suggest these actors are making any persistent contribution to the conflict, even in a major war between two nations with a long history of cyber warfare. Their role and capacity in future nation-state conflicts should not be confounded with state hacking or political `hacktivism'. Our diverse, separately collected datasets all point to a narrative that notable attention was temporarily drawn to Russia and Ukraine rather than other countries. Neither the engagement on \hackforums nor Telegram, or the outbreak of defacements and DDoS attacks, was long-lasting -- presumably as participants just lost interest, despite their choice of targets being influenced by the war for a while. This is in line with other work suggesting that boredom is an important factor in people leaving cybercrime~\cite{collier2021cybercrime}.

We do not dispute claims about the prevalence of state-sponsored attacks such as malware and phishing~\cite{fogofwar,microsoftreport}, but rather provide additional perspectives on the role of low-level cybercrime actors. Some cybercrime-related activities may indeed contribute to the war effort. Leaks, especially of high-profile datasets gathered from Russian public services, have consistently made headlines. They may or may not be connected to civilians, hacktivists, state actors, or other groups. Much as with ransomware, their low numbers and vast disparities in impact make them far less cross-comparable. Our findings fit a more general pattern in the cybercrime ecosystem increasingly characterised by an entrepreneurial, service-based economy which is becoming alienated from traditional hacker culture's concerns with technical learning and dissent~\cite{anderson2021}. Committed, persistent hacktivists appear to be separate from low-level crime communities whose interest seems to have been fleeting and easily diverted by trending news. They did indeed briefly get involved in the war effort using off-the-shelf exploit tools, but their role on the `hard' digital frontline remains rather limited. These are best seen as actions in the theatre of protest, `soft power' and solidarity. In a `cyberwar', efforts should be prioritised against high-profile and state-sponsored threats instead of such low-level cybercrime actors, even though they may cause immediately noticeable effects.

\section*{Acknowledgments}
We thank the anonymous reviewers. This work is supported by the European Research Council (ERC) under the European Union's Horizon 2020 research and innovation programme, grant No 949127.

\bibliographystyle{acm-style} \bibliography{main}
\appendix

\section{Ethical Considerations} \label{appendix:ethical-issues}
This work is presented objectively to minimise risk to researchers. The collection, sharing, and analysis of web defacement, amplified DDoS attacks, and Telegram chats have been formally approved by the Department of Computer Science \& Technology's Research Ethics Committee. We do not attempt to gather private data; only publicly accessible data are collected. A 2022 US court ruling suggests scraping public data is legal~\cite{webscrapinglegal}; our scraper does not overload websites. The amplified DDoS attack honeypots absorb attack packets without relaying them, thus reducing harm to victims.

Studying an ongoing conflict may harm individuals whose attacks are reported, while researchers might face retaliation from attackers due to leaking insights into their activities and community. To avoid potential harm, our experiments operate ethically and collectively, only presenting aggregated findings without identifying individuals. We did not ask for consent from Telegram users or web defacers when using scraped data, as sending thousands of messages would be impractical. We believe they are aware that content posted online will be publicly visible. This approach accords with the British Society of Criminology's Statement on Ethics~\cite{britishethics}.

\section{Data Licensing} \label{appendix:data-licensing}
We have robust procedures and long experience in licensing our data in various jurisdictions. Our quantitative data, analysis scripts, and scrapers are available for academic researchers under a license agreement with the \ccc~to prevent misuse and to ensure the data will be treated ethically, as access to sensitive data might risk researchers and the actors involved~\cite{doerfler2021m,warford2022sok}.

\section{Determining Attack Geolocation} \label{appendix:determining-geolocation}
Accurately mapping IP addresses to countries is challenging as IP geolocation is not always stable and trustworthy~\cite{gouel2021ip}; providers prefer locating servers in countries with cheap hosting~\cite{weinberg2018catch}. Attack geolocation can thus be determined differently e.g., \zoneh says an IP is in Germany, while \zonexsec detects Singapore and \defacerid cannot tell. IP geolocation may be more reliable at the country level~\cite{cozar2022reliability}, but this is only part of the truth as websites are nowadays commonly hosted on content delivery networks (CDNs), where original IP addresses are hidden and the geolocation is of the CDNs.

For example, a `.ru' website is supposed to be Russian, but it might be physically hosted in Vietnam, operated by a person living in Hong Kong, while proxied through Cloudflare with an IP address in the US. Relying on only one aspect might be risky, as both IP and domain can lie. We use data fusion to enhance the accuracy, prioritising: (1) top-level domain; (2) IP geolocation at collection time (MaxMind GeoIP2\footnote{~GeoIP2 is freely accessible at https://maxmind.com/. It offers both free and paid licenses, with the paid one being slightly more accurate and up-to-date. It claims to provide over 99.8\% country-level and over 60\% city-level accuracy, yet that varies from country to country e.g., 79\% for Russia and 65\% for Ukraine, within a 250km radius.} for web defacements, and a database we maintain based on Regional Internet Registry data for DDoS attacks; and (3) geolocation of the AS hosting the IP address. If a website's IP address belongs to a CDN, its geolocation is determined solely by ccTLD, as any geolocation of IP address or ASN will be unreliable.

The top three CDNs are Cloudflare, Amazon Web Services, and Akamai, serving around 89\% of customers~\cite{cdnmarketshare}. We ignore the trivial market shares of their competitors but count \ddosguard as it is Russian-based, which may affect the measurement. We expect the four to cover over 90\% of customers. We found \CSnDefacementHostedOnCDN\% of defacements are hosted on these CDNs by \MSnCDNPrefixesTotal~prefixes as of the writing date: \MSnPrefixesCloudflare~of Cloudflare; \MSnPrefixesAmazonWS~of Amazon Webservice; \MSnPrefixesAkamai~of Akamai; and \MSnPrefixesDDoSGuard~of \ddosguard (these prefixes and AS number mappings are collected on Hurricane Electric Internet Services). For defacements, we prefer ccTLDs over IP geolocation as attackers often target sites by massively scanning domain ccTLDs, such as `.ru' and `.ua', rather than checking whether IP addresses are hosted in those countries.

Accurate measurement of frequent ccTLDs used in Russia and Ukraine is complex; many Ukrainian firms use Russian services and vice versa. The domain most frequently used in Russia is `.ru'~\cite{commondomainsofru}. A similar report for Ukraine is unavailable, but we believe incorporating ccTLDs with IP and AS geolocation is reasonable as choosing targets based on ccTLDs is a common way used by low-lever actors.

\section{Defacement Submission Process} \label{appendix:web-defacement-submission-process}
The defacement submission is mostly automatic: users specify a `notifier', team, defaced URL, vulnerability types, and hacking incentives. At that point, a record is made with details of the compromised system, its IP address and location, and a snapshot of the defaced page (often consisting of messages that may include political and ideological propaganda~\cite{banerjee2021using}). Messages can be hidden using identical font colours as the background, but are detectable through HTML. New reports are kept away from the dashboard until being verified by staff or bots. Although `notifier' can be arbitrarily entered, defacers are incentivised to use consistent handles to cultivate fame and reputation; we thus consider `notifier' to be reliable enough to differentiate them. The defacement snapshots, including messages left, are highly reliable as they are captured at reporting time.

\section{Website Defacement Collection} \label{appendix:web-defacement-collection}
Data completeness and reliability are critical for longitudinal measurement. Scraping \textit{complete} defacement archives at scale, especially \zoneh, is non-trivial, and was not guaranteed in prior work. Some purchased the data~\cite{maggi2018investigating}, but it is not sustainable and is ethically questionable. This is challenging as (1) \zoneh uses \captcha to prevent bots, (2) its dashboard sets a limit of 50 pages where older data is hidden, and (3) on-hold records may not appear promptly, leading to potential misses. The only way to get a complete scrape is by iterating all submission IDs, with IDs of valid and invalid reports often mixed. This issue, along with IP blacklisting and bot prevention, generates a non-trivial workload. We responded by (1) developing an efficient text \captcha solver for \zoneh using image-processing techniques, (2) routing our scraper through multiple proxies, and (3) iterating all submission IDs in turn. Raw data is stored in a database to avoid unnecessary future requests.

Five most trusted archives are included; an active one \defaceridsmall (since February 2016) is excluded as (1) the valid submission volume during the period is small (less than 27k); (2) unclear staff verification, no validity sanitisation on submission, no validity signal in defaced pages (in fact, over half of these have been deemed invalid by the archive); (3) defaced snapshots and defacers' messages are missing; and (4) victim geolocation is mostly lacking; determining it after the fact is problematic as sites could have been relocated.

\begin{algorithm}[t]
\caption{Semi-automatic website defacement validation}
\label{algo:defacement-validation}
\small
\begin{algorithmic}[1]
    \Procedure{validate\_defacements}{}
        \ForEach {$a \in$ verifiedDefacements()}\Comment{verified by archives}
            \State a.status $\gets 0$ \Comment{originally validated}
        \EndFor
        \ForEach {$a \in$ filteredDefacements()}\Comment{filtered by terms}
            \State a.status $\gets 1$ \Comment{automatically validated}
        \EndFor
        \State $P \gets $ pendingGroups()\Comment{groups of pending attacks}
        \State $V \gets $ verifiedGroups()\Comment{groups of verified attacks}
        \ForEach {$p \in\mathcal P$}
            \State $T \gets \{\}$
            \ForEach {$v \in\mathcal V$}
                \State $d \gets$ levenshtein($p, v$) \Comment{similarity with verified ones}
                \State $T \gets$ topSimilar($d, T$) \Comment{extract top similar ones}
            \EndFor  
            \State showSimilarDefacements($T$)\Comment{to assist the annotator}
            \State $s \gets$ annotation()\Comment{annotate the validity}
            \ForEach {$a \in p$} \Comment{each defacement in this group}
                \State a.status $\gets s$\Comment{update validation status}
            \EndFor
            \If{isValidated($s$)}\Comment{if it is manually validated}
                \State $V \gets V \cap p$  \Comment{add to validated groups}
            \EndIf
        \EndFor
    \EndProcedure
  \end{algorithmic}
\end{algorithm}

\begin{table*}[t]
\centering
\small
\caption{Significance levels of the impact on daily defacements and defacers targeting Russia, Ukraine, and top countries.}
\setlength{\tabcolsep}{0.52em}
\begin{tabular}{l|rrrrr|rrrrr}
\toprule
\multirow{2}{*}{Country} & \multicolumn{5}{c|}{Tests for the number of web defacements per day} & \multicolumn{5}{c}{Tests for the number of web defacers per day}\\
\cmidrule{2-11}
& ANOVA / Kruskal-Wallis & \textbf{$\langle E_1, E_2 \rangle$} & \textbf{$\langle E_1, E_3 \rangle$} & \textbf{$\langle E_2, E_3 \rangle$} & \textbf{$\eta^2$} & ANOVA / Kruskal-Wallis & \textbf{$\langle E_1, E_2 \rangle$} & \textbf{$\langle E_1, E_3 \rangle$} & \textbf{$\langle E_2, E_3 \rangle$} & \textbf{$\eta^2$}\\
\midrule
Russia & $H(\CSnDefacementsRUSignificanceTestDoF) = \CSnDefacementsRUSignificanceTestStat, \CSnDefacementsRUSignificanceTestPValue$ & $\CSnDefacementsRUPostHocPValueONETWO$ & \textcolor{lightgray}{$\CSnDefacementsRUPostHocPValueONETHREE$} &  $\CSnDefacementsRUPostHocPValueTWOTHREE$ & \CSnDefacementsRUEffectSize & $H(\CSnDefacersRUSignificanceTestDoF) = \CSnDefacersRUSignificanceTestStat, \CSnDefacersRUSignificanceTestPValue$ & $\CSnDefacersRUPostHocPValueONETWO$ & \textcolor{lightgray}{$\CSnDefacersRUPostHocPValueONETHREE$} & $\CSnDefacersRUPostHocPValueTWOTHREE$ & \CSnDefacersRUEffectSize\\
Ukraine & $H(\CSnDefacementsUASignificanceTestDoF) = \CSnDefacementsUASignificanceTestStat, \CSnDefacementsUASignificanceTestPValue$ & $\CSnDefacementsUAPostHocPValueONETWO$ & \textcolor{lightgray}{$\CSnDefacementsUAPostHocPValueONETHREE$} &  $\CSnDefacementsUAPostHocPValueTWOTHREE$ & \CSnDefacementsUAEffectSize & $H(\CSnDefacersUASignificanceTestDoF) = \CSnDefacersUASignificanceTestStat, \CSnDefacersUASignificanceTestPValue$ & $\CSnDefacersUAPostHocPValueONETWO$ & \textcolor{lightgray}{$\CSnDefacersUAPostHocPValueONETHREE$} &  $\CSnDefacersUAPostHocPValueTWOTHREE$ & \CSnDefacersUAEffectSize\\
\arrayrulecolor{black!50}
\midrule
The US & $H(\CSnDefacementsUSSignificanceTestDoF) = \CSnDefacementsUSSignificanceTestStat, \CSnDefacementsUSSignificanceTestPValue$ & $\CSnDefacementsUSPostHocPValueONETWO$ & \textcolor{lightgray}{$\CSnDefacementsUSPostHocPValueONETHREE$} &  $\CSnDefacementsUSPostHocPValueTWOTHREE$ & \CSnDefacementsUSEffectSize & $H(\CSnDefacersUSSignificanceTestDoF) = \CSnDefacersUSSignificanceTestStat, \CSnDefacersUSSignificanceTestPValue$ & $\CSnDefacersUSPostHocPValueONETWO$ & \textcolor{lightgray}{$\CSnDefacersUSPostHocPValueONETHREE$} &  $\CSnDefacersUSPostHocPValueTWOTHREE$ & \CSnDefacersUSEffectSize\\
Brazil & \textcolor{lightgray}{$H(\CSnDefacementsBRSignificanceTestDoF) = \CSnDefacementsBRSignificanceTestStat, \CSnDefacementsBRSignificanceTestPValue$} & \textcolor{lightgray}{$\CSnDefacementsBRPostHocPValueONETWO$} & \textcolor{lightgray}{$\CSnDefacementsBRPostHocPValueONETHREE$} & \textcolor{lightgray}{ $\CSnDefacementsBRPostHocPValueTWOTHREE$} & \textcolor{lightgray}{\CSnDefacementsBREffectSize} & $H(\CSnDefacersBRSignificanceTestDoF) = \CSnDefacersBRSignificanceTestStat, \CSnDefacersBRSignificanceTestPValue$ & \textcolor{lightgray}{$\CSnDefacersBRPostHocPValueONETWO$} &  $\CSnDefacersBRPostHocPValueONETHREE$ &  $\CSnDefacersBRPostHocPValueTWOTHREE$ & \textcolor{lightgray}{\CSnDefacersBREffectSize}\\
Germany & \textcolor{lightgray}{$H(\CSnDefacementsDESignificanceTestDoF) = \CSnDefacementsDESignificanceTestStat, \CSnDefacementsDESignificanceTestPValue$} & \textcolor{lightgray}{$\CSnDefacementsDEPostHocPValueONETWO$} & \textcolor{lightgray}{$\CSnDefacementsDEPostHocPValueONETHREE$} & \textcolor{lightgray}{$\CSnDefacementsDEPostHocPValueTWOTHREE$} & \textcolor{lightgray}{\CSnDefacementsDEEffectSize} & $F(\CSnDefacersDESignificanceTestDoFBetween, \CSnDefacersDESignificanceTestDoFWithin) = \CSnDefacersDESignificanceTestStat, \CSnDefacersDESignificanceTestPValue$ & \textcolor{lightgray}{$\CSnDefacersDEPostHocPValueONETWO$} & \textcolor{lightgray}{$\CSnDefacersDEPostHocPValueONETHREE$} & \textcolor{lightgray}{$\CSnDefacersDEPostHocPValueTWOTHREE$} & \textcolor{lightgray}{\CSnDefacersDEEffectSize}\\
India & \textcolor{lightgray}{$H(\CSnDefacementsINSignificanceTestDoF) = \CSnDefacementsINSignificanceTestStat, \CSnDefacementsINSignificanceTestPValue$} & \textcolor{lightgray}{$\CSnDefacementsINPostHocPValueONETWO$} & \textcolor{lightgray}{$\CSnDefacementsINPostHocPValueONETHREE$} & \textcolor{lightgray}{$\CSnDefacementsINPostHocPValueTWOTHREE$} & \textcolor{lightgray}{\CSnDefacementsINEffectSize} & \textcolor{lightgray}{$H(\CSnDefacersINSignificanceTestDoF) = \CSnDefacersINSignificanceTestStat, \CSnDefacersINSignificanceTestPValue$} & \textcolor{lightgray}{$\CSnDefacersINPostHocPValueONETWO$} & \textcolor{lightgray}{$\CSnDefacersINPostHocPValueONETHREE$} & \textcolor{lightgray}{$\CSnDefacersINPostHocPValueTWOTHREE$} & \textcolor{lightgray}{\CSnDefacersINEffectSize}\\
Indonesia & $H(\CSnDefacementsIDSignificanceTestDoF) = \CSnDefacementsIDSignificanceTestStat, \CSnDefacementsIDSignificanceTestPValue$ & $\CSnDefacementsIDPostHocPValueONETWO$ & \textcolor{lightgray}{$\CSnDefacementsIDPostHocPValueONETHREE$} &  $\CSnDefacementsIDPostHocPValueTWOTHREE$ & \textcolor{lightgray}{\CSnDefacementsIDEffectSize} & $H(\CSnDefacersIDSignificanceTestDoF) = \CSnDefacersIDSignificanceTestStat, \CSnDefacersIDSignificanceTestPValue$ & $\CSnDefacersIDPostHocPValueONETWO$ & \textcolor{lightgray}{$\CSnDefacersIDPostHocPValueONETHREE$} &  $\CSnDefacersIDPostHocPValueTWOTHREE$ & \CSnDefacersIDEffectSize\\
Canada & $H(\CSnDefacementsCASignificanceTestDoF) = \CSnDefacementsCASignificanceTestStat, \CSnDefacementsCASignificanceTestPValue$ & \textcolor{lightgray}{$\CSnDefacementsCAPostHocPValueONETWO$} & \textcolor{lightgray}{$\CSnDefacementsCAPostHocPValueONETHREE$} &  $\CSnDefacementsCAPostHocPValueTWOTHREE$ & \textcolor{lightgray}{\CSnDefacementsCAEffectSize} & $H(\CSnDefacersCASignificanceTestDoF) = \CSnDefacersCASignificanceTestStat, \CSnDefacersCASignificanceTestPValue$ & \textcolor{lightgray}{$\CSnDefacersCAPostHocPValueONETWO$} & \textcolor{lightgray}{$\CSnDefacersCAPostHocPValueONETHREE$} &  $\CSnDefacersCAPostHocPValueTWOTHREE$ & \textcolor{lightgray}{\CSnDefacersCAEffectSize}\\
Turkey & $H(\CSnDefacementsTRSignificanceTestDoF) = \CSnDefacementsTRSignificanceTestStat, \CSnDefacementsTRSignificanceTestPValue$ & \textcolor{lightgray}{$\CSnDefacementsTRPostHocPValueONETWO$} &  $\CSnDefacementsTRPostHocPValueONETHREE$ &  $\CSnDefacementsTRPostHocPValueTWOTHREE$ & \CSnDefacementsTREffectSize & $H(\CSnDefacersTRSignificanceTestDoF) = \CSnDefacersTRSignificanceTestStat, \CSnDefacersTRSignificanceTestPValue$ & \textcolor{lightgray}{$\CSnDefacersTRPostHocPValueONETWO$} &  $\CSnDefacersTRPostHocPValueONETHREE$ &  $\CSnDefacersTRPostHocPValueTWOTHREE$ & \CSnDefacersTREffectSize\\
Singapore & \textcolor{lightgray}{$H(\CSnDefacementsSGSignificanceTestDoF) = \CSnDefacementsSGSignificanceTestStat, \CSnDefacementsSGSignificanceTestPValue$} & \textcolor{lightgray}{$\CSnDefacementsSGPostHocPValueONETWO$} & \textcolor{lightgray}{$\CSnDefacementsSGPostHocPValueONETHREE$} & \textcolor{lightgray}{ $\CSnDefacementsSGPostHocPValueTWOTHREE$} & \textcolor{lightgray}{\CSnDefacementsSGEffectSize} & \textcolor{lightgray}{$H(\CSnDefacersSGSignificanceTestDoF) = \CSnDefacersSGSignificanceTestStat, \CSnDefacersSGSignificanceTestPValue$} & \textcolor{lightgray}{$\CSnDefacersSGPostHocPValueONETWO$} &  $\CSnDefacersSGPostHocPValueONETHREE$ & \textcolor{lightgray}{ $\CSnDefacersSGPostHocPValueTWOTHREE$} & \textcolor{lightgray}{\CSnDefacersSGEffectSize}\\
\arrayrulecolor{black}
\midrule
\end{tabular}
\label{tab:statistical-significance-web-defacements}
\end{table*}

\begin{table*}[t]
\centering
\small
\caption{Significance levels of the impact on daily DDoS attacks and victims targeting Russia, Ukraine, and top countries.}
\setlength{\tabcolsep}{0.48em}
\begin{tabular}{l|rrrrr|rrrrr}
\toprule
\multirow{2}{*}{Country} & \multicolumn{5}{c|}{Tests for the number of DDoS attacks per day} & \multicolumn{5}{c}{Tests for the number of DDoS victims per day}\\
\cmidrule{2-11}
& ANOVA / Kruskal-Wallis & \textbf{$\langle E_1, E_2 \rangle$} & \textbf{$\langle E_1, E_3 \rangle$} & \textbf{$\langle E_2, E_3 \rangle$} & \textbf{$\eta^2$} & ANOVA / Kruskal-Wallis & \textbf{$\langle E_1, E_2 \rangle$} & \textbf{$\langle E_1, E_3 \rangle$} & \textbf{$\langle E_2, E_3 \rangle$} & \textbf{$\eta^2$}\\
\midrule
Russia & $H(\CSnDDoSAttacksRUSignificanceTestDoF) = \CSnDDoSAttacksRUSignificanceTestStat, \CSnDDoSAttacksRUSignificanceTestPValue$ & $\CSnDDoSAttacksRUPostHocPValueONETWO$ & $\CSnDDoSAttacksRUPostHocPValueONETHREE$ &  $\CSnDDoSAttacksRUPostHocPValueTWOTHREE$ & \CSnDDoSAttacksRUEffectSize & $H(\CSnDDoSVictimsRUSignificanceTestDoF) = \CSnDDoSVictimsRUSignificanceTestStat, \CSnDDoSVictimsRUSignificanceTestPValue$ & $\CSnDDoSVictimsRUPostHocPValueONETWO$ &  $\CSnDDoSVictimsRUPostHocPValueONETHREE$ &  $\CSnDDoSVictimsRUPostHocPValueTWOTHREE$ & \CSnDDoSVictimsRUEffectSize\\
Ukraine & $H(\CSnDDoSAttacksUASignificanceTestDoF) = \CSnDDoSAttacksUASignificanceTestStat, \CSnDDoSAttacksUASignificanceTestPValue$ & $\CSnDDoSAttacksUAPostHocPValueONETWO$ & \textcolor{lightgray}{$\CSnDDoSAttacksUAPostHocPValueONETHREE$} &  $\CSnDDoSAttacksUAPostHocPValueTWOTHREE$ & \CSnDDoSAttacksUAEffectSize & $H(\CSnDDoSVictimsUASignificanceTestDoF) = \CSnDDoSVictimsUASignificanceTestStat, \CSnDDoSVictimsUASignificanceTestPValue$ & $\CSnDDoSVictimsUAPostHocPValueONETWO$ & \textcolor{lightgray}{$\CSnDDoSVictimsUAPostHocPValueONETHREE$} &  $\CSnDDoSVictimsUAPostHocPValueTWOTHREE$ & \CSnDDoSVictimsUAEffectSize\\
\arrayrulecolor{black!50}
\midrule
The US & $H(\CSnDDoSAttacksUSSignificanceTestDoF) = \CSnDDoSAttacksUSSignificanceTestStat, \CSnDDoSAttacksUSSignificanceTestPValue$ & \textcolor{lightgray}{$\CSnDDoSAttacksUSPostHocPValueONETWO$} &  $\CSnDDoSAttacksUSPostHocPValueONETHREE$ & \textcolor{lightgray}{ $\CSnDDoSAttacksUSPostHocPValueTWOTHREE$} & \textcolor{lightgray}{\CSnDDoSAttacksUSEffectSize} & \textcolor{lightgray}{$H(\CSnDDoSVictimsUSSignificanceTestDoF) = \CSnDDoSVictimsUSSignificanceTestStat, \CSnDDoSVictimsUSSignificanceTestPValue$} & \textcolor{lightgray}{$\CSnDDoSVictimsUSPostHocPValueONETWO$} &  $\CSnDDoSVictimsUSPostHocPValueONETHREE$ & \textcolor{lightgray}{$\CSnDDoSVictimsUSPostHocPValueTWOTHREE$} & \textcolor{lightgray}{\CSnDDoSVictimsUSEffectSize}\\
Brazil & $H(\CSnDDoSAttacksBRSignificanceTestDoF) = \CSnDDoSAttacksBRSignificanceTestStat, \CSnDDoSAttacksBRSignificanceTestPValue$ & $\CSnDDoSAttacksBRPostHocPValueONETWO$ & \textcolor{lightgray}{$\CSnDDoSAttacksBRPostHocPValueONETHREE$} &  $\CSnDDoSAttacksBRPostHocPValueTWOTHREE$ & \textcolor{lightgray}{\CSnDDoSAttacksBREffectSize} & $H(\CSnDDoSVictimsBRSignificanceTestDoF) = \CSnDDoSVictimsBRSignificanceTestStat, \CSnDDoSVictimsBRSignificanceTestPValue$ & $\CSnDDoSVictimsBRPostHocPValueONETWO$ &  $\CSnDDoSVictimsBRPostHocPValueONETHREE$ & $\CSnDDoSVictimsBRPostHocPValueTWOTHREE$ & \CSnDDoSVictimsBREffectSize\\
Germany & $H(\CSnDDoSAttacksDESignificanceTestDoF) = \CSnDDoSAttacksDESignificanceTestStat, \CSnDDoSAttacksDESignificanceTestPValue$ & $\CSnDDoSAttacksDEPostHocPValueONETWO$ &  $\CSnDDoSAttacksDEPostHocPValueONETHREE$ & \textcolor{lightgray}{ $\CSnDDoSAttacksDEPostHocPValueTWOTHREE$} & \textcolor{lightgray}{\CSnDDoSAttacksDEEffectSize} & $H(\CSnDDoSVictimsDESignificanceTestDoF) = \CSnDDoSVictimsDESignificanceTestStat, \CSnDDoSVictimsDESignificanceTestPValue$ & $\CSnDDoSVictimsDEPostHocPValueONETWO$ &  $\CSnDDoSVictimsDEPostHocPValueONETHREE$ & \textcolor{lightgray}{ $\CSnDDoSVictimsDEPostHocPValueTWOTHREE$} & \CSnDDoSVictimsDEEffectSize\\
Bangladesh & \textcolor{lightgray}{$H(\CSnDDoSAttacksBDSignificanceTestDoF) = \CSnDDoSAttacksBDSignificanceTestStat, \CSnDDoSAttacksBDSignificanceTestPValue$} & $\CSnDDoSAttacksBDPostHocPValueONETWO$ & \textcolor{lightgray}{$\CSnDDoSAttacksBDPostHocPValueONETHREE$} & \textcolor{lightgray}{$\CSnDDoSAttacksBDPostHocPValueTWOTHREE$} & \textcolor{lightgray}{\CSnDDoSAttacksBDEffectSize} & \textcolor{lightgray}{$H(\CSnDDoSVictimsBDSignificanceTestDoF) = \CSnDDoSVictimsBDSignificanceTestStat, \CSnDDoSVictimsBDSignificanceTestPValue$} & $\CSnDDoSVictimsBDPostHocPValueONETWO$ & \textcolor{lightgray}{$\CSnDDoSVictimsBDPostHocPValueONETHREE$} & \textcolor{lightgray}{$\CSnDDoSVictimsBDPostHocPValueTWOTHREE$} & \textcolor{lightgray}{\CSnDDoSVictimsBDEffectSize}\\
China & $H(\CSnDDoSAttacksCNSignificanceTestDoF) = \CSnDDoSAttacksCNSignificanceTestStat, \CSnDDoSAttacksCNSignificanceTestPValue$ & $\CSnDDoSAttacksCNPostHocPValueONETWO$ & $\CSnDDoSAttacksCNPostHocPValueONETHREE$ & $\CSnDDoSAttacksCNPostHocPValueTWOTHREE$ & \CSnDDoSAttacksCNEffectSize & $H(\CSnDDoSVictimsCNSignificanceTestDoF) = \CSnDDoSVictimsCNSignificanceTestStat, \CSnDDoSVictimsCNSignificanceTestPValue$ & $\CSnDDoSVictimsCNPostHocPValueONETWO$ & $\CSnDDoSVictimsCNPostHocPValueONETHREE$ & \textcolor{lightgray}{$\CSnDDoSVictimsCNPostHocPValueTWOTHREE$} & \CSnDDoSVictimsCNEffectSize\\
France & $H(\CSnDDoSAttacksFRSignificanceTestDoF) = \CSnDDoSAttacksFRSignificanceTestStat, \CSnDDoSAttacksFRSignificanceTestPValue$ & \textcolor{lightgray}{$\CSnDDoSAttacksFRPostHocPValueONETWO$} &  $\CSnDDoSAttacksFRPostHocPValueONETHREE$ &  $\CSnDDoSAttacksFRPostHocPValueTWOTHREE$ & \CSnDDoSAttacksFREffectSize & $H(\CSnDDoSVictimsFRSignificanceTestDoF) = \CSnDDoSVictimsFRSignificanceTestStat, \CSnDDoSVictimsFRSignificanceTestPValue$ & \textcolor{lightgray}{$\CSnDDoSVictimsFRPostHocPValueONETWO$} &  $\CSnDDoSVictimsFRPostHocPValueONETHREE$ & \textcolor{lightgray}{ $\CSnDDoSVictimsFRPostHocPValueTWOTHREE$} & \textcolor{lightgray}{\CSnDDoSVictimsFREffectSize}\\
UK & $H(\CSnDDoSAttacksGBSignificanceTestDoF) = \CSnDDoSAttacksGBSignificanceTestStat, \CSnDDoSAttacksGBSignificanceTestPValue$ & \textcolor{lightgray}{$\CSnDDoSAttacksGBPostHocPValueONETWO$} &  $\CSnDDoSAttacksGBPostHocPValueONETHREE$ &  $\CSnDDoSAttacksGBPostHocPValueTWOTHREE$ & \CSnDDoSAttacksGBEffectSize & $H(\CSnDDoSVictimsGBSignificanceTestDoF) = \CSnDDoSVictimsGBSignificanceTestStat, \CSnDDoSVictimsGBSignificanceTestPValue$ & \textcolor{lightgray}{$\CSnDDoSVictimsGBPostHocPValueONETWO$} &  $\CSnDDoSVictimsGBPostHocPValueONETHREE$ & \textcolor{lightgray}{ $\CSnDDoSVictimsGBPostHocPValueTWOTHREE$} & \textcolor{lightgray}{\CSnDDoSVictimsGBEffectSize}\\
Poland & \textcolor{lightgray}{$H(\CSnDDoSAttacksPLSignificanceTestDoF) = \CSnDDoSAttacksPLSignificanceTestStat, \CSnDDoSAttacksPLSignificanceTestPValue$} & \textcolor{lightgray}{$\CSnDDoSAttacksPLPostHocPValueONETWO$} & \textcolor{lightgray}{$\CSnDDoSAttacksPLPostHocPValueONETHREE$} & \textcolor{lightgray}{$\CSnDDoSAttacksPLPostHocPValueTWOTHREE$} & \textcolor{lightgray}{\CSnDDoSAttacksPLEffectSize} & \textcolor{lightgray}{$H(\CSnDDoSVictimsPLSignificanceTestDoF) = \CSnDDoSVictimsPLSignificanceTestStat, \CSnDDoSVictimsPLSignificanceTestPValue$} & \textcolor{lightgray}{$\CSnDDoSVictimsPLPostHocPValueONETWO$} & \textcolor{lightgray}{$\CSnDDoSVictimsPLPostHocPValueONETHREE$} & \textcolor{lightgray}{$\CSnDDoSVictimsPLPostHocPValueTWOTHREE$} & \textcolor{lightgray}{\CSnDDoSVictimsPLEffectSize}\\
\arrayrulecolor{black}
\midrule
\end{tabular}
\label{tab:statistical-significance-ddos-attacks}
\end{table*}

\section{Validating On-hold Defacements} \label{appendix:validating-defacement}
How defacement reports are validated is not clearly stated. \zoneh reports are kept on hold until being manually verified by staff; \zonexsec, \defacerpro, and \haxorid use automatic validation, insisting messages left on defaced pages contain keywords linked to hacking activities (e.g., \textit{`Hacked by ABC'}); while it is unclear for \ownzyou. Defacers may game the system by putting comments on blogs, or submitting search queries (e.g., ?search=`Hacked by ABC'), which occasionally get through automatic sanitisation but our further validation excludes them. Manual staff review on \zoneh may be slow, while automatic verification of the others is error-prone. Unverified records may be kept on hold forever, leading to incomplete data. Consequently, collecting only defacements shown in the dashboard is inadequate, making a complete dataset challenging to gather. To enhance data completeness and reliability, a semi-automatic validation is performed to check if on-hold reports are in fact valid.

Our strategy is shown in Algorithm~\ref{algo:defacement-validation}. First, reports verified by archives are considered valid. Second, messages on the defaced pages of on-hold submissions are used to decide the validity, as defacers often leave signatures for reputation e.g., `\textit{Hacked by CoolHacker}'. If messages include defacer handles and specifically contain common hacking terms: \textit{`hacked by'}, \textit{`h4ck3d by'}, \textit{`h4cked by'}, \textit{`p4wn3d by'}, \textit{`pwn3d by'}, \textit{`pwnd by'}, \textit{`pwned by'}, \textit{`pwndz by'}, \textit{`owned by'}, \textit{`own3d by'}, \textit{`touched by'}, and \textit{`kissed by'}, we consider them to be valid e.g., a message \textit{`This website was hacked, contact me t.me/coolhacker'} posted by a notifier \textit{`CoolHacker'} is considered valid. This method is looser than an exact comparison with \textit{`Hacked by CoolHacker'}, but is still highly accurate; 100 randomly checked samples were all correct. Third, the remaining submissions are manually validated by looking for defacer signatures; some are obvious but some are complicated. Candidates are grouped by normalised handles and messages (redundant spaces removed), and then for each, the 10 most similar validated defacements are suggested to the annotator. Levenshtein distance is used to estimate the similarity between two messages, which is helpful as messages are often slightly modified from templates. If no message is found (instead images, iframes, or javascript), or leftover signatures cannot be spotted, a browser opens the defaced page to assist annotators. 

This assistance effectively reduces the annotation effort. One challenge is defaced pages redirecting to another page that can be modified dynamically: when they are down, the submission may point to a non-existent site, but a careful check could reveal evidence of defacers. We also consider a site is touched if its content is unchanged but the page title is modified to indicate hacking activities. We ignore cases that lack evidence to ensure those flagged valid are indeed valid. We do not use complex machine learning techniques as message texts contain lots of noise; given a small number of samples (around 10k), machine learning is not more effective than a rule-based approach. Sometimes, defacements appeared to be already verified at the collection time, but became invalid afterwards; we re-validated them months after the initial collection to make sure their status had been finalised by the archives.

\section{Unifying Defacements and Defacers} \label{appendix:unifying-defacement-defacer}
We hashed then unified defacements across all archives based on reporting dates, original defacer handles, root victim domains, and messages left on the defaced page. Hashing reporting dates may cause repeat counts if defacers resubmit to other archives after a few days, but excluding them may lead to missing defacements of the same URL on different days due to repeat victimisation. We also unified defacers across all archives, as users tend to pick similar pseudonyms on different platforms~\cite{goga2013exploiting}. As the unification needs to be accurate and the number of unique defacers does not exceed a few thousand, machine learning is not appropriate. We instead used a semi-automatic approach combining automated handle similarity analysis with manual review. As typos may occur (e.g., missing characters, character orders, case sensitivity), similar handle pairs are first extracted using Levenshtein distance, which is set to not exceed 25\% of the length. Ten messages left by defacers in each pair are then sampled to assist the annotation. A pair of handles are unified under a single nickname if their messages are closed enough, assessed based on handle appearances, message semantics, stylometry, synonyms, typos, team, nationality, languages, and handle rarity (rare ones such as `cj2ks' are more likely to be used by a single person, while common ones such as `glory' are more likely to be shared by multiple individuals~\cite{liu2013s}). Messages left are diverse: many defacers leave identical messages on different archives, while some are relatively similar, some are distinct and contain the defacer names, and some consist of their phone numbers. We only confirm when having sufficient evidence, uncertain pairs are left unmatched.

\section{Statistical Significance of Impact} \label{appendix:significance-impact-levels}
Tables~\ref{tab:statistical-significance-web-defacements} and~\ref{tab:statistical-significance-ddos-attacks} detail the statistical significance levels of the impact on defacement and DDoS attack counts over the three eras, using One-way ANOVA or Kruskal-Wallis tests. Post-hoc analysis Tukey-Kramer is used for ANOVA, while Dunn's is used for Kruskal-Wallis.

\end{document}